\documentclass[iop,twocolumn,apj,numberedappendix,appendixfloats]{emulateapj}
\usepackage{apjfonts,color,soul}
\usepackage[pagebackref=false,letterpaper=true,colorlinks=true,citecolor=blue,linkcolor=blue,breaklinks=true,bookmarks=true]{hyperref}

\usepackage{mathrsfs}
\usepackage{epsfig}
\usepackage{graphicx}
\usepackage{mathtools}
\usepackage{color}
\usepackage{url}
\usepackage{hyperref}
\usepackage{footnote}

\newcommand{\ly}{\rm{Ly}\alpha}
\newcommand{\lz}{\rm{low}\hbox{-}z}
\newcommand{\eor}{\rm{EoR}}
\newcommand{\dgl}{\rm{DGL}}
\newcommand{\ihl}{\rm{IHL}}
\newcommand{\shot}{\rm{shot}}

\newcommand{\ppp}{\rm{PopIII}}
\newcommand{\pp}{\rm{PopII}}
\newcommand{\hii}{\rm{HII}}
\newcommand{\hi}{\rm{HI}}

\newcommand{\neb}{\rm{neb}}
\newcommand{\igm}{\rm{IGM}}
\newcommand{\ff}{\rm{ff}}
\newcommand{\fb}{\rm{fb}}
\newcommand{\phph}{2\hbox{-}\rm{photon}}
\newcommand{\um}{\mu\rm{m}}

\begin{document}
\title{Multi-Component Decomposition of Cosmic Infrared Background Fluctuations}

\author{Chang Feng\altaffilmark{1,2}, Asantha Cooray\altaffilmark{2},  Jamie Bock\altaffilmark{3,4}, Tzu-Ching Chang\altaffilmark{3,4,5}, Olivier Dor\'e\altaffilmark{3,4}, Mario G. Santos\altaffilmark{6,7,10}, Marta B. Silva\altaffilmark{9},  Michael Zemcov\altaffilmark{8,4}}\email{changf@illinois.edu}

\altaffiltext{1}{Department of Physics, University of Illinois at Urbana-Champaign, 1110 W Green St, Urbana, IL, 61801, USA}
\altaffiltext{2}{Department of Physics and Astronomy, University of California, Irvine, CA 92697, USA }
\altaffiltext{3}{California Institute of Technology, 1200 E. California Blvd., Pasadena, CA 91125}
\altaffiltext{4}{Jet Propulsion Laboratory, California Institute of Technology, Pasadena CA 91109, USA}
\altaffiltext{5}{Institute of Astronomy and Astrophysics, Academia Sinica, Roosevelt Rd, Taipei 10617, Taiwan}
\altaffiltext{6}{Department of Physics and Astronomy, University of the Western Cape, Cape Town, South Africa}
\altaffiltext{7}{SKA South Africa, The Park, Park Road, Pinelands, 7405, South Africa}
\altaffiltext{8}{Rochester Institute of Technology, Rochester, NY\ 14623}
\altaffiltext{9}{Kapteyn Astronomical Institute, University of Groningen, Landleven 12, 9747AD Groningen, the Netherlands}
\altaffiltext{10}{Instituto de Astrofisica e Ciencias do Espaco, Universidade de Lisboa, OAL, Tapada da Ajuda, PT1349-018 Lisboa, Portugal}

\begin{abstract}
The near-infrared background between 0.5 $\mu$m to 2 $\mu$m contains a wealth of information related to radiative processes in the universe. Infrared background anisotropies encode the redshift-weighted total emission over cosmic history, including any spatially diffuse and extended contributions. The anisotropy power spectrum is dominated by undetected galaxies at small angular scales and diffuse background of Galactic emission at large angular scales. In addition to these known sources, the infrared background also arises from intra-halo light (IHL) at $z < 3$ associated with tidally-stripped stars during galaxy mergers. Moreover, it contains information on the very first galaxies from the epoch of reionization (EoR). The EoR signal has a spectral energy distribution (SED) that goes to zero near optical wavelengths due to Lyman absorption, while other signals have spectra that vary smoothly with frequency. Due to differences in SEDs and spatial clustering, these components may be separated in a multi-wavelength-fluctuation experiment. To study the extent to which EoR fluctuations can be separated in the presence of IHL, extra-galactic and Galactic foregrounds, we develop a maximum likelihood technique that incorporates a full covariance matrix among all the frequencies at different angular scales. We apply this technique to simulated deep imaging data over a 2$\times$100 deg$^2$ sky area from 0.75 $\mu$m to 5 $\mu$m in 9 bands and find that such a ``frequency tomography'' can successfully reconstruct both the amplitude and spectral shape for representative EoR, IHL and the foreground signals.
\end{abstract}

\maketitle

\newpage
\section{Introduction}
The optical and infrared background traces nucleosynthesis in stars and radiation from black holes throughout cosmic history. In addition to sources in our Galaxy, the absolute infrared background intensity is composed of diffuse sources of emission~\citep{2002MNRAS.336.1082S},  such as intra-halo light (IHL)~\citep{2012arXiv1210.6031C, 2014Sci...346..732Z} and faint galaxies present during the epoch of reionization (EoR). Instead of absolute intensity, which can be easily contaminated by Galactic emission, the near-infrared background may be studied using spatial fluctuations or anisotropies~\citep{2016RSOS....350555C}. These have been measured in broad continuum bands by a number of groups~\citep{2011ApJ...742..124M, 2005Natur.438...45K, 2012arXiv1210.6031C, 2012ApJ...753...63K, 2015NatCo...6E7945M, 2014Sci...346..732Z}. The fluctuation amplitude, which is robustly consistent across these measurements, exceeds that expected from the large-scale clustering of known galaxy populations~\citep{2012ApJ...752..113H}. 

The EoR signal is associated with the first collapsed objects that formed and produced energetic ultraviolet (UV) photons that reionized the surrounding hydrogen gas. In theoretical models~\citep{2012ApJ...756...92C, 2012ApJ...750...20F}, the expected amplitude of this component is several orders of magnitude below the level from faint galaxies in the more nearby universe. The emission of the ionizing photons during reionization is expected to peak between 0.9 and 1.1 $\mu$m today, if the implied optical depth is consistent with Planck result~\citep{2016A&A...596A.108P} and the reionization occurred around $z\sim$ 7 to 9. Also, it is damped quickly shortward of $\sim0.8\, \mu \rm{m}$~\citep{2004ApJ...606..683S, 2003MNRAS.339..973S} due to Lyman absorption. While the amplitude is small, component separation is possible through the unique spectral dependence afforded by the
Lyman drop-out signature, similar to the Lyman drop-out signature used to identify bright galaxies present at $z > 6$ during reionization. Therefore, spatial fluctuations of the infrared (IR) background centered around 1 $\mu$m provide a way to discriminate the signal generated by galaxies present during reionization from those at lower redshifts, based on the strength of the drop-out signature in the fluctuations measured in different bands. 

In a recent study~\citep{2015NatCo...6E7945M}, a model with different IR fluctuations was constrained by multi-wavelength power spectra in five bands conducted with the Hubble Space Telescope (HST)/Advanced Camera for Surveys (ACS) and Wide Field Camera 3 (WFC3) data from the Cosmic Assembly Near-IR Deep Extragalactic Legacy Survey (CANDELS). Instead of only using auto-correlations of the HST measurements that are largely limited by the number of broad bands~\citep{2015NatCo...6E7945M}, in this work, we develop a novel component separation approach with full covariance information, including cross correlations of fluctuations across multiple frequency bands, to extract both spectral and spatial information for various astrophysical components. The cross correlations of spatial fluctuations between different wavelengths can break degeneracies with model parameters of other components, thereby providing crucial information about the EoR. This type of component separation will be crucial for upcoming measurements that will be performed with more than five spectral channels, such as the Cosmic Infrared Background ExpeRiment-2 (CIBER-2)\footnotemark[1]\footnotetext[1]{https://cosmology.caltech.edu/projects/ciber}~\citep{2014Sci...346..732Z,2016SPIE.9904E..4JS} and the proposed all-sky Spectro-Photometer for the History of the Universe, Epoch of Reionization, and Ices Explorer (SPHEREx) \footnotemark[2]\footnotetext[2]{http://spherex.caltech.edu/}, both of which can provide ideal data sets to measure the IR background fluctuations and study the EoR~\citep{2016arXiv160607039D, 2014arXiv1412.4872D}. In this work, we focus on the SPHEREx experiment, which has one observing straitening that will produce two surveys of different depths: an all-sky shallow survey and two deep surveys near the ecliptic poles. SPHEREx provides measurements in a total of 96 bands from 0.75 to 5 $\mu$m, and we only focus on fluctuations measured by combining those narrow-band images to nine broad-band images -- 0.8, 0.9, 1.025, 1.2, 1.5, 2.0, 2.65, 3.5, 4.5 $\mu$\rm{m} -- over a 2$\times$100 deg$^2$ sky area from the deep survey. Even with the compressed bands, the component separation scheme starts to become computationally challenging as the total number of parameters grows very fast when all the cosmic infrared background (CIB) sources have to be modeled at $N$ broad bands. 

This paper is organized as follows: In Section 2 we summarize various CIB components and the methodology related to component separation is introduced in Section 3;
in Section 4 we apply the component-separation method to the simulated SPHEREx data;
we conclude with a summary in Section 5. Unless otherwise noted we assume a cosmological model consistent with latest Planck CMB measurements~\citep{2016A&A...594A..13P}.\\

\section{Component power spectra}

In this Section, we outline how our theory models relate to various components of the angular power spectrum of IR background anisotropies. Our calculations follow existing models in the literature. In the following, we will discuss detailed models that will be used for the component separation.

\subsection{Angular power spectrum of each CIB component}

We begin with the halo model formalism~\citep{2002PhR...372....1C}, which can be used to calculate the power spectrum using a combination of 1-halo:
\begin{eqnarray}
C_l^{1h,XY}&=&\int dz\frac{d\chi}{dz}\Big(\frac{a}{\chi}\Big)^2\nonumber\\
&&\times\int dM n(M,z)X_l(k,M,z)Y_l(k,M,z)\label{1halo}
\end{eqnarray}

and 2-halo terms of clustering as
\begin{eqnarray}
C_l^{2h,XY}&=&\int dz\frac{d\chi}{dz}\Big(\frac{a}{\chi}\Big)^2P_{\rm{lin}}(k,z)\nonumber\\
&&\times\Big[\int dMb(M,z)n(M,z)\tilde X_l(k,M,z)\Big]\nonumber\\
&&\times\Big[\int dMb(M,z)n(M,z)\tilde Y_l(k,M,z)\Big],\label{2halo}
\end{eqnarray}
respectively. Here $X$ or $Y$ = \{\rm{IHL}, \lz, \rm{EoR}\}, $a$ is the scale factor, $\chi$ is comoving distance, $b(M,z)$ and $n(M,z)$ are bias and mass functions for halo mass $M$ at redshift $z$, and $P_{\rm lin}$ is the linear matter power spectrum at scale $k$ which is $\ell/\chi$ when the Limber approximation is assumed. Now we derive all the shape factors $X(k,M,z)$, $\tilde X(k,M,z)$, $Y(k,M,z)$ and $\tilde Y(k,M,z)$ in the next subsections.

\subsection{Galaxies and IHL}

Low-redshift galaxies are known CIB emitters. To model their contributions to the CIB, we first introduce the occupation numbers of central and satellite galaxies for a dark matter halo of mass $M$:
\begin{equation}
N_c(M)=\frac{1}{2}\Big[1+{\rm erf}\Big(\frac{\log_{10}M-\log_{10}{M_{{\rm min}}}}{\sigma_M}\Big)\Big]\label{nc}
\end{equation}
and 
\begin{equation}
N_s(M)=\frac{1}{2}\Big[1+{\rm erf}\Big(\frac{\log_{10}M-\log_{10}{2M_{{\rm min}}}}{\sigma_M}\Big)\Big]\Big(\frac{M}{M_s}\Big)^{\alpha_s}.\label{ns}
\end{equation}
Here $M_{\rm{min}}$ = $10^{9}M_{\odot}$, $\sigma_{M}$ = 0.2, $M_s$ = $5\times10^{10}M_{\odot}$ and $\alpha_s$ = 1~\citep{2011ApJ...736...59Z}. For $\lz$ galaxy occupation number, we have $N_g(M)$ = $N_c(M)+N_s(M)$ and $N_g(N_g-1)$ = $2N_cN_s+N_s^2$. The template luminosity function is parametrized as~\citep{2012ApJ...752..113H}:
\begin{eqnarray}
\Phi_{\nu}(M_b)&=&0.4\ln(10)\phi^{\ast}_{\nu}[10^{0.4(M^{\ast}_{b,\nu}-M_b)}]^{\alpha_{\nu}+1}\nonumber\\
&&\times e^{-10^{0.4(M^{\ast}_{b,\nu}-M_b)}},
\end{eqnarray}
where 
$M^{\ast}_{b,\nu}(z)$ = $M^{\ast}_{b,0}-2.5\log_{10}[1+(z-z_0)^q]$, $M_b$ is absolute magnitude, $\phi^{\ast}_{\nu}(z)$ = $\phi^{\ast}_0e^{-p(z-z_0)}$,  and $\alpha_{\nu}(z)$ = $\alpha_0(z/z_0)^r$. The values assumed for each parameter are given in Table~\ref{lowzLF} ~\citep{2012ApJ...752..113H}.

From the luminosity function, the number of galaxies per redshift is given by:
\begin{equation}
\frac{dN_{\nu}(m_b)}{dz}=\Phi_{\nu}(m_b,z)\frac{dV}{dzd\Omega},
\end{equation}
and the flux production rate is given by:
\begin{equation}
\Big(\frac{dF}{dz}\Big)_{\nu}=\nu\int^{\infty}_{m_{\rm{b,lim}}}dm_b10^{-0.4(m_b-23.9)}\frac{dN_{\nu}(m_b,z)}{dz}.\label{lowzflux}
\end{equation} 
Here $(m_b-M_b)$ is the distance modulus and $dV/dz/d\Omega$ is the comoving volume per solid angle.
As discussed in~\cite{2012ApJ...752..113H}, we take the lower bound of a limiting magnitude $m_{\rm{b,lim}}=22$, which indicates that a mask is applied to the deep survey images and bright point sources are removed. This number is consistent with the sensitivity requirement studied by~\cite{2016arXiv160607039D} for a deep survey in ecliptic poles with SPHEREx. A small limiting magnitude corresponds to less masking of the point sources. To test if deep masking is required, we run the component separation against a no-masking scheme with $m_{\rm{b,lim}}=10$ and find that the reconstructions of different components are not affected. We will discuss this test in detail in Section 4.
 
The masking will determine emissivity of the unresolved $\lz$ galaxy because it is related to the flux production rate via:
\begin{equation}
\nu j^{\lz}_{\nu}(z)=\frac{1}{a}\frac{dz}{d\chi}\Big(\frac{dF}{dz}\Big)_{\nu}.
\end{equation}

With this emissivity function, we can construct the halo model for the \lz\ galaxy, i.e., $X(k, M, z)$ or $Y(k, M, z)$ = $\nu j^{\lz}_{\nu}(z)\sqrt{2N_cN_su(k,M,z)+N_s^2u^2(k,M,z)}/{\bar n_g}$ in Eq. (\ref{1halo}), and  $\tilde X(k, M, z)$ or $\tilde Y(k, M, z)$ = $\nu j^{\lz}_{\nu}(z)N_g/{\bar n_g}u(k,M,z)$ in Eq. (\ref{2halo}). Here $u(k,M,z)$ is the Fourier transform of the Navarro-Frenk-White (NFW) profile ~\citep{1996ApJ...462..563N} and $\bar n_g$ is the averaged number density of galaxies.

During galaxy-merging processes, stars could be stripped from their host galaxies. These stars are too faint to be detected individually at other wavelengths, but can be detected collectively at infrared wavelengths. The infrared emission from these stars is called intrahalo light. To model the intrahalo light, we use an empirical luminosity function~\citep{2012arXiv1210.6031C}:
\begin{equation}
\bar L^{\rm{IHL}}_{\nu}(M,z)=f_{\rm{IHL}}L_{2.2}(M)(1+z)^{p_{\rm{IHL}}}F_{\lambda}^{\rm{IHL}}.
\end{equation}
Here the luminosity fraction due to IHL at halo with mass $M$ is $f_{\rm{IHL}}(M)$ = $\beta_{\rm{IHL}}(M/M_0)^{\alpha_{\rm{IHL}}}$, the total luminosity function of halo with mass $M$ at redshift $z=0$ is $L_{2.2}(M)$ = $5.64\times10^{12}h^{-2}_{70}(M/(2.7\times10^{14}h^{-1}_{70}M_{\odot}))^{0.72}L_{\odot}$
 and $h_{70}=100h/70$. The spectral energy distribution (SED)  $F_{\lambda}^{\rm{IHL}}$ of IHL is generated from old stellar populations and normalized at 2.2 $\um$ and $z=0$. The parameters for the fiducial model are $\log_{10}{\beta_{\rm{IHL}}}$ = -2.04, $\alpha_{\rm{IHL}}$ = 0.1, $p_{\rm{IHL}}$ = -1.05 and $M_0$ = $10^{12}M_{\odot}$. The IHL shape factor $X(k,M,z)$ (= $Y(k,M,z)$ = $\tilde X(k,M,z)$ = $\tilde Y(k,M,z)$) = $\nu \bar L^{\ihl}_{\nu}u(k,M,z)/(4\pi)$. We should note that this IHL model can fit the recent CIB observations but is not derived from first principles, moreover, there are other possible interpretations of the IHL signal~\citep{2013ApJ...769...68C, 2013MNRAS.433.1556Y}. In the future, dedicated numerical simulations and high sensitivity CIB measurements at multiple wavelengths will be able to test different IHL models. 

\begin{table}
\caption{Best-fit evolution parameters for the luminosity function of known galaxy populations. The values are taken from a previous study~~\citep{2012ApJ...752..113H}.}
\begin{center}\resizebox{\linewidth}{!}{%
\begin{tabular}{ c|c|c|c|c|c|c|c} 
 \hline
$\lambda$&$z_{\rm{max}}$&$M^{\ast}_{b0}$&$q$&$\phi^{\ast}_{0}[10^{-3}\rm{Mpc}^{-3}]$&$p$&$\alpha_{0}$&$r$\\
\hline
 0.89-0.92$\mu$m&2.9&-22.86&0.4&2.55&0.4&-1&0.06\\
 1.24-1.27$\mu$m&3.2&-23.04&0.4&2.21&0.6&-1&0.035\\
 3.6$\mu$m&0.7&-22.40&0.2&3.29&0.8&-1&0.035\\
 4.5$\mu$m&0.7&-21.84&0.3&3.29&0.8&-1&0.035\\
\hline
\end{tabular}}\label{lowzLF}
\end{center}
\end{table}

\begin{figure*}
\includegraphics[width=8cm, height=6cm]{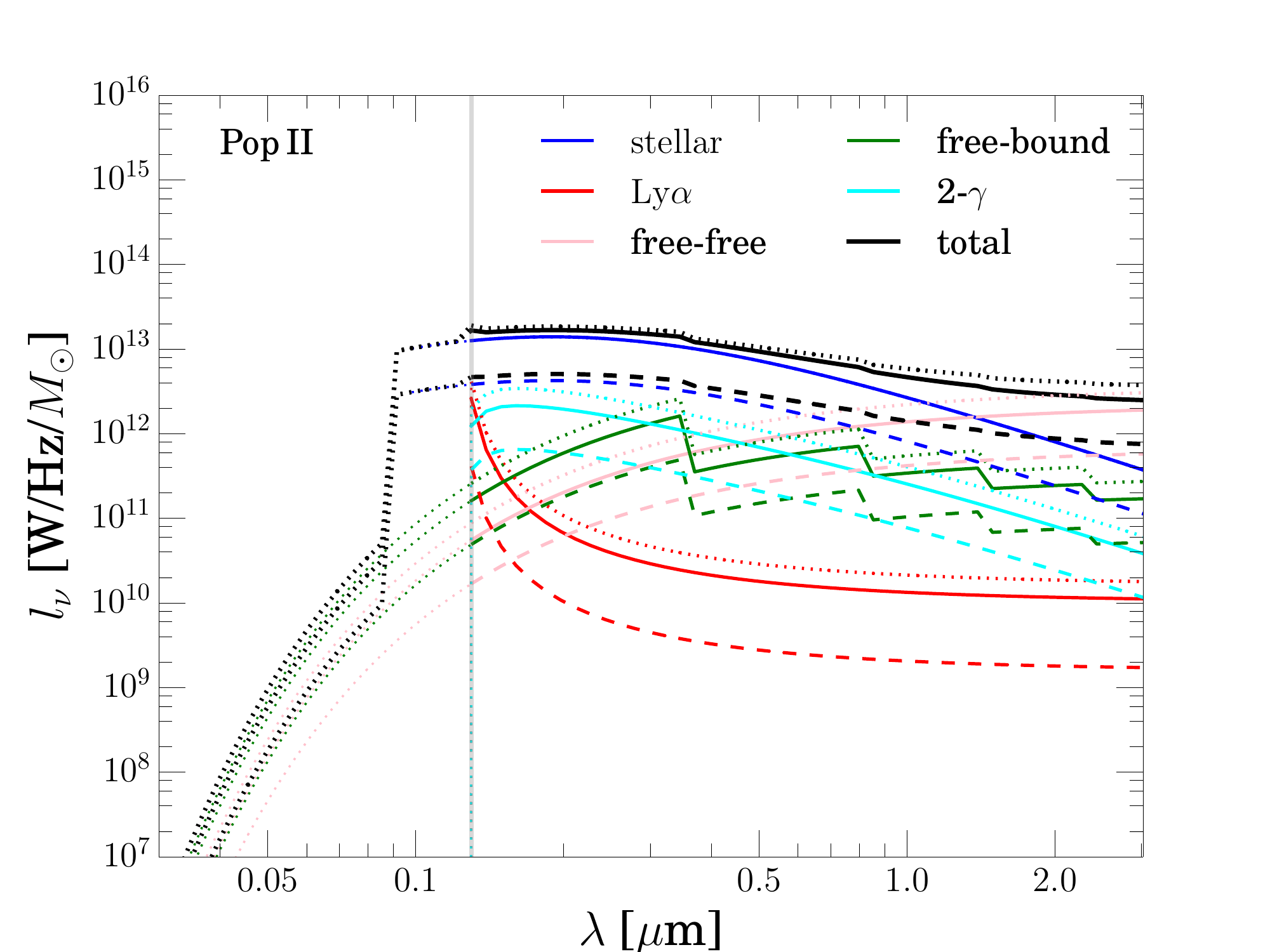}
\includegraphics[width=8cm, height=6cm]{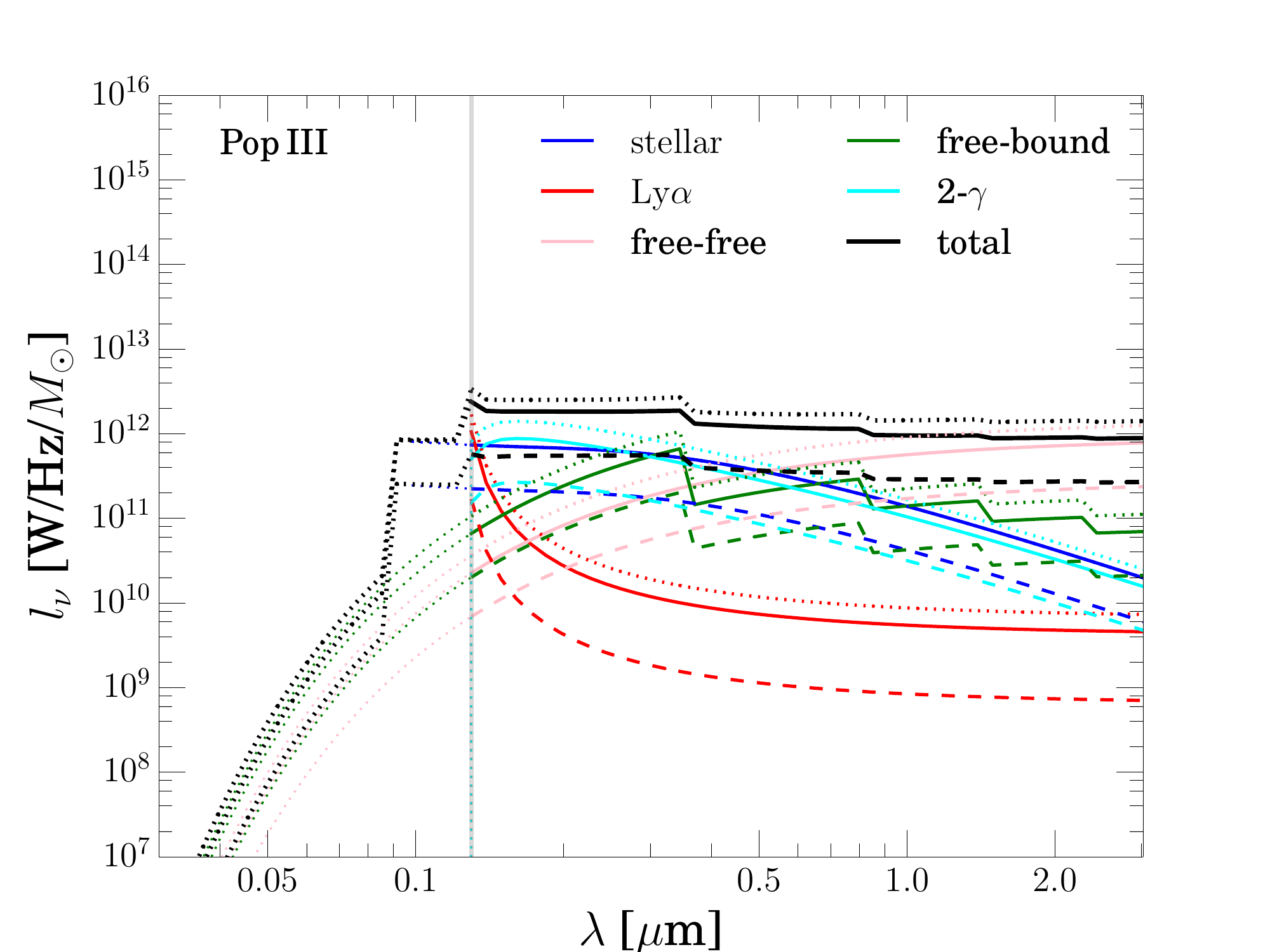}
\caption{Luminosity mass density $l^{(i)}_{\nu}$ vs rest-frame wavelength. Three models are shown in different line styles: solid lines -- ($z=10$, $f_{\rm esc}=0.5$, $f_{\ast}=0.03$); dotted lines -- ($z=10$, $f_{\rm esc}=0.2$, $f_{\ast}=0.06$); dashed lines -- ($z=6$, $f_{\rm esc}=0.5$, $f_{\ast}=0.03$). All the components, including stellar, Lyman-$\alpha$, free-free, free-bound, two-photon emission are shown for each model. The vertical gray line denotes the wavelength of Ly$\alpha$ at rest-frame at $z=0$. 
The cut off above the Ly$\alpha$ frequency is due to absorption by dust and scattering by neutral hydrogen gas.}\label{lpopIIandIII}
\end{figure*}

\subsection{EoR signal}

Besides infrared emissions from known galaxies and stripped stars at low redshifts, the first stars and galaxies during the epoch of reionization (EoR) can emit energetic photons which are redshifted to the infrared wavelengths as we observe today. This infrared signature encodes rich information about EoR so it is of great importance. We follow the standard model~\citep{2012ApJ...756...92C} to calculate the emissivity functions of Pop III and Pop II stars at high redshift. We use the empirical fitting formula from simulations to determine the stellar initial mass function $f(M_{\ast})$, the intrinsic bolometric luminosity $L_{\ast}^{\rm{bol}}$, the effective temperature $T^{\rm{eff}}_{\ast}$, the main-sequence life time $\tau_{\ast}$ and the time-averaged hydrogen photoionization rate $Q_{\rm{HI}}$ from~\cite{2012ApJ...756...92C}. Based on these fitting functions, we can calculate averaged stellar mass $\langle M_{\ast}\rangle$, main sequence life-time $\langle \tau_{\ast}\rangle$ and hydrogen-reionization rate $\langle Q_{\rm{HI}}\rangle$ from:
\begin{equation}
\langle X_{\ast}\rangle=\int dM_{\ast}X_{\ast}(M_{\ast})f(M_{\ast}),
\end{equation}
where $X$ refers to $M_{\ast}$, $\tau_{\ast}$ and $Q_{\rm{HI}}$. We use a population function to combine both Pop II and Pop III stars and the population function is $f_{\rm{P}}$ = $[1+{\rm{erf}}((z-z_t)/\sigma_t)]/2$. Here $z_t$ = 10 and $\sigma_t$ = 0.5. The EoR emissivity:
\begin{equation}
\bar j^{\rm{EoR}}_{\nu}(z)=(1-f_{\rm{P}})\bar j_{\nu}^{\pp}+f_P\bar j_{\nu}^{\ppp}
\end{equation}
is related to the volume-averaged luminosity mass density $l^{(i)}_{\nu}$ by:
\begin{equation}
\bar j^{(i)}_{\nu}(z)=\frac{1}{4\pi}l^{(i)}_{\nu}\langle\tau^{(i)}_{\ast}\rangle\psi(z),
\end{equation}
where:
\begin{equation}
\psi(z)=f_{\ast}\frac{\Omega_b}{\Omega_m}\frac{d}{dt}\int_{M_{\rm{min}}}dM M n(M,z)
\end{equation}
is the star formation density. Here $n(M, z)$ is the halo mass function and $M_{\rm min}=10^8M_{\odot}$.

The functions of luminosity mass density in the nebulae and IGM are given by a few components and they are $l^{(i)}_{\neb}$ = $l_{\ast}^{(i)}$ + ($1-f_{\rm{esc}}$)($l_{\ly}^{(i)}+l_{\ff}^{(i)}+l_{\fb}^{(i)}+l_{\phph}^{(i)}$) and 
$L^{(i)}_{\igm}$ = $f_{\rm{esc}}$($L_{\ly}^{(i)}+L_{\ff}^{(i)}+L_{\fb}^{(i)}+L_{\phph}^{(i)}$), respectively. Here the superscript ${}^{(i)}$ refers to $\pp$ and $\ppp$, and the subscripts ``$\ast$'', ``$\ly$'', ``$\ff$'', ``$\fb$'' and ``$\phph$'' refer to stellar, $\ly$, free-free, free-bound, and 2-photon emissions. The lower case $l$ denotes the luminosity from nebulae and the upper case $L$ refers to that of IGM. The difference mostly comes from the emission volume. In Fig. \ref{lpopIIandIII}, we show luminosity mass densities versus the rest-frame wavelength. As an approximation to the Lyman-$\alpha$ absorption, we apply a sharp cutoff at Lyman-$\alpha$ wavelength to the luminosity mass densities. In reality, the cutoff is essentially a soft decrease blueward of the Lyman-$\alpha$ frequency, and here we consider an idealized case that approximates the transition as a step function. We note that there is a considerable decrease in flux above the Lyman-$\alpha$ frequency due to dust absorption and scattering in the IGM which could greatly enhance the CIB emission during the EoR. For the calculations, we set the gas temperature to be $3\times 10^4$ \rm{K}. The local electron and \hii\ densities in nebulae region are set to $n_e$ = $n_{\hii}$ = $10^4\,\rm{cm}^{-3}$. In the IGM region, the electron and \hii\,densities depend on the ionizing fraction $x_{\hii}$ that is solved from:
\begin{equation}
\frac{dx_{\hii}}{dt}=\frac{f_{\rm{esc}}\psi(z)q(z)}{\bar n_H(z)}-\frac{x_{\hii}}{\bar t_{\rm{rec}}}.\label{xii}
\end{equation}
The initial condition for this equation is $x_{\hii}(z=\infty)=0$. The derived optical depth today is consistent with the latest Planck result $\tau=0.058\pm0.012$~\citep{2016A&A...596A.108P}.

In this equation:
\begin{equation}
q(z)=(1-f_{\rm{P}})\frac{\langle Q_{\hi}^{\pp}\rangle}{\langle M_{\ast}^{\pp}\rangle}\langle \tau_{\ast}^{\pp}\rangle+f_{\rm{P}}\frac{\langle Q_{\hi}^{\ppp}\rangle}{\langle M_{\ast}^{\ppp}\rangle}\langle \tau_{\ast}^{\ppp}\rangle
\end{equation}
and $t_{\rm{rec}}^{-1}$ = $C_{\hii}(z)\alpha_{\rm{B}}^{\rm{rec}}\bar n_{\rm{H}}(1+Y/4X)$. We use the fitting formula for the clumping factor of the ionized hydrogen $C_{\hii}(z)$ and $C_{\hii}(z)$ = $2.9[(1+z)/6]^{-1.1}$~\citep{2011arXiv1108.3334S}, $\alpha_{\rm{B}}^{\rm{rec}}$ is the hydrogen case B recombination coefficient given in~\cite{2012ApJ...756...92C}, and $\bar n_{\rm H}$ is the averaged baryon density. The mass fractions of hydrogen and helium are $X=0.75$ and $Y=0.25$. 
More detailed discussions of the emissions of stellar, free-free, free-bound, 2-photon and $\ly$ can be found in~\cite{2012ApJ...756...92C}. For the occupation numbers of the first stars and galaxies, we use Eqs. (\ref{nc}) and (\ref{ns}) and assume $M_s=15 M_{\rm{min}}$, $\sigma_{M}=0.3$ and $\alpha_s=1.5$. For the angular power spectrum of the EoR, the shape factors are $X(k, M, z)$ or $Y(k, M, z)$ = $\nu {\bar j}^{\eor}_{\nu}(z)\sqrt{2N_cN_su(k,M,z)+N_s^2u^2(k,M,z)}/{\bar n_g}$ and  $\tilde X(k, M, z)$ or $\tilde Y(k, M, z)$ = $\nu {\bar j}^{\eor}_{\nu}(z)N_g/{\bar n_g}u(k,M,z)$ for Eqs. (\ref{1halo}) and (\ref{2halo}), respectively.

\subsection{Diffuse Galactic Light}
Starlight scattered by interstellar dust forms the diffuse Galactic light (DGL) which at certain frequencies dominates the CIB contribution on large angular scales~\citep{2012ApJ...744..129B}. Here we assume that the power spectrum of the DGL follows a simple power law $C_{\ell}^{\dgl}=A\ell^{-3}$ where the amplitude $A$ has been measured by CIBER over a broad range of wavelengths~\citep{2015ApJ...806...69A}.

\subsection{Shot Noise}
Also, there is a critical flux threshold above which galaxy clustering can be measured. Below the flux threshold, the discrete nature of the sources leads to a shot noise component which is the dominant component in the CIB emission at very small angular scales. The shot noise power spectrum is flat so it is simply $C_{\ell}^{\shot}$ = const, and the amplitude varies from band to band. 

\begin{figure}
\includegraphics[width=8cm, height=6cm]{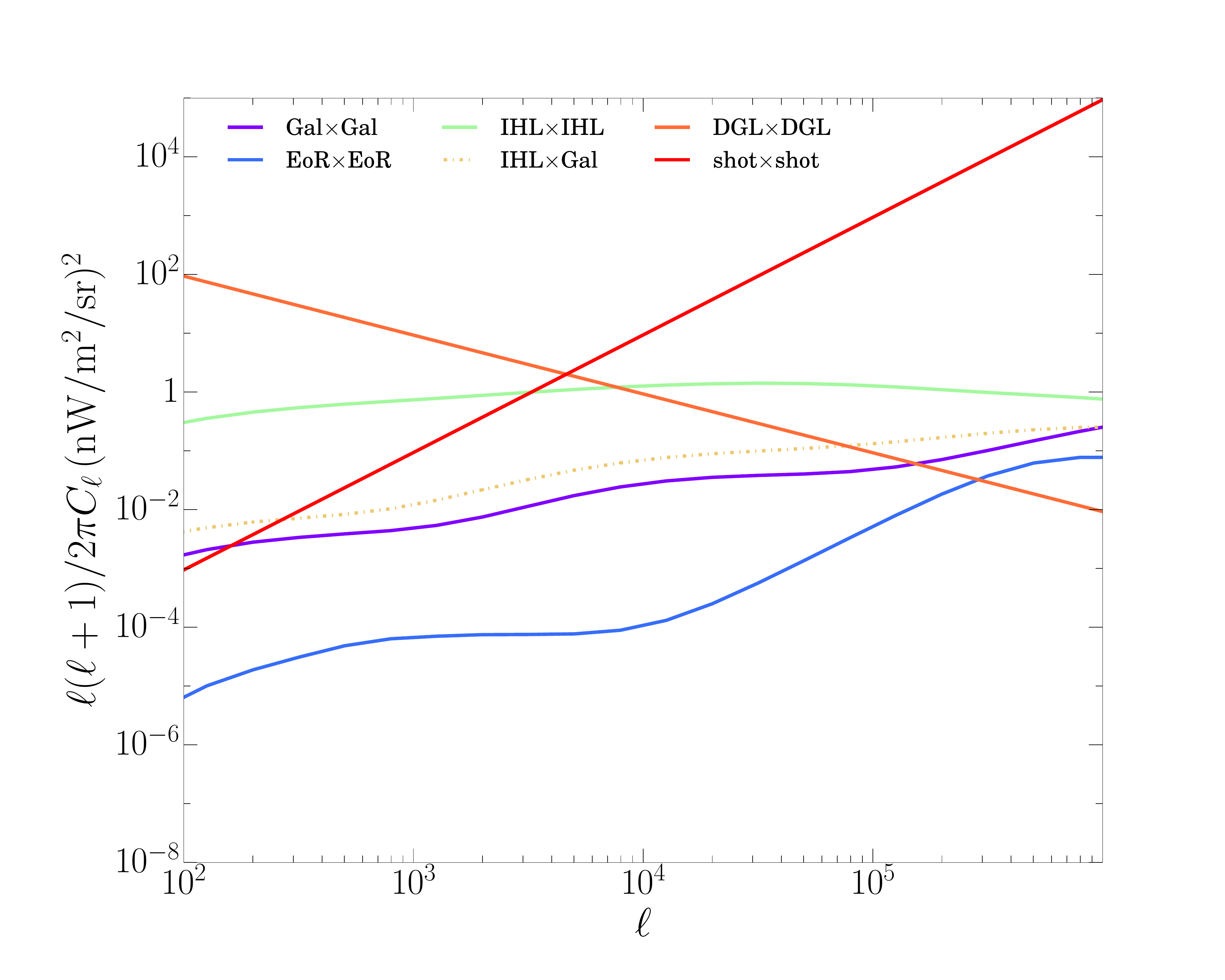}
\caption{A representative plot with all the CIB components at 1.02$\mu$\rm{m}.}\label{rep1um}
\end{figure}

In Fig. \ref{rep1um}, we show a representative plot for all the CIB components discussed in this section. The components are calculated for an observed wavelength of 1.02 $\mu$\rm{m}. It is seen from this figure that the CIB is dominated by the DGL and shot noise at large and small angular scales, respectively. Deep masking can effectively reduce the contamination from unresolved galaxies on the \eor\ signal, but the cross correlation between the bright IHL and the faint \lz\ galaxy population after masking is still significant as the orange dot-dashed line shows. Thus, to detect the very faint EoR signature from a single wavelength would be very challenging.

The instrumental noise should be included in the CIB model as well. We use the noise model estimated from both raw detector sensitivity and two-year optimal scan pattern on the deep fields for the analysis. We extensively investigated the optimal scan strategy for a 24-month integration time, and chose the best scanning strategy that results in the most uniform and largest number counts per pixel on the deep fields. Based on this optimal scan pattern, we estimated both optimistic and pessimistic noise levels for SPHEREx. 

\section{Multi-Frequency Tomography}

The measured infrared (IR) background fluctuations can originate from IHL, known galaxies at low redshift (\lz), diffuse Galactic light (DGL), and most importantly, the first stars and galaxies (\eor). 
We write down the surface brightness ($I_{\nu}$) IR model as a sum of different brightness templates for each component, whose amplitude $A_{\nu}$ will be determined by the component separation method:
\begin{eqnarray}
I_{\nu}&=&A^{\ihl}_{\nu} I^{\rm{IHL}}_{\nu}+A^{\eor}_{\nu}I^{\rm{EoR}}_{\nu}+A^{\lz}_{\nu} I^{\lz}_{\nu}\nonumber\\
&+&A^{\dgl}_{\nu}I^{\rm{DGL}}_{\nu}+A^{\shot}_{\nu}I^{\rm{Shot}}_{\nu}.\label{cib}
\end{eqnarray}
Here the shot noise term includes contributions from all the CIB components. 

The IR power spectrum that is calculated from IR background anisotropy maps at two frequencies $\nu_i$ and $\nu_j$ can be written as
\begin{eqnarray}
C_{\ell}^{(\nu_i\nu_j)}&=&A^{\ihl}_iA^{\ihl}_jC_{\ell}^{\ihl_i-\ihl_j}+A^{\eor}_iA^{\eor}_jC_{\ell}^{\eor_i-\eor_j}\nonumber\\
&+&A^{\lz}_iA^{\lz}_jC_{\ell}^{\lz_i-\lz_j}+A^{\ihl}_iA^{\eor}_jC_{\ell}^{\ihl_i-\eor_j}\nonumber\\
&+&A^{\ihl}_iA^{\lz}_jC_{\ell}^{\ihl_i-\lz_j}+A^{\eor}_iA^{\lz}_jC_{\ell}^{\eor_i-\lz_j}+(i\leftrightarrow j)\nonumber\\
&+&(A\ast A)^{\dgl}_{ij}C_{\ell}^{\dgl_i-\dgl_j}+(A\ast A)^{\shot}_{ij}C_{\ell}^{\shot_i-\shot_j},\label{component_ps}
\end{eqnarray}
where the $C_{\ell}$ are calculated from Eqs. (\ref{1halo}) and (\ref{2halo}).
We allow both the amplitude and slope of each component to vary, i.e., $A^{\ihl}_i$, $A^{\lz}_i$ and $A^{\eor}_{i}$ are independent parameters at each broad band. There is a cutoff below $0.9\mu\rm{m}$ in the spectrum of EoR fluctuations due to the Lyman-$\alpha$ dropout feature in the rest-frame so we set the amplitude below $0.9\mu\rm{m}$ to be 0. 
The amplitudes ${(A\ast A)}^{\dgl}_{ij}$ and $(A\ast A)^{\shot}_{ij}$ for both the DGL and shot noise can be precisely fitted from very low- and high-$\ell$ power spectra, respectively. We constrain $3N-1$ parameters for $N(N+1)/2$ power spectrum measurements. 

\begin{figure*}
\includegraphics[trim=70 0 0 0, width=10cm, height=8.5cm]{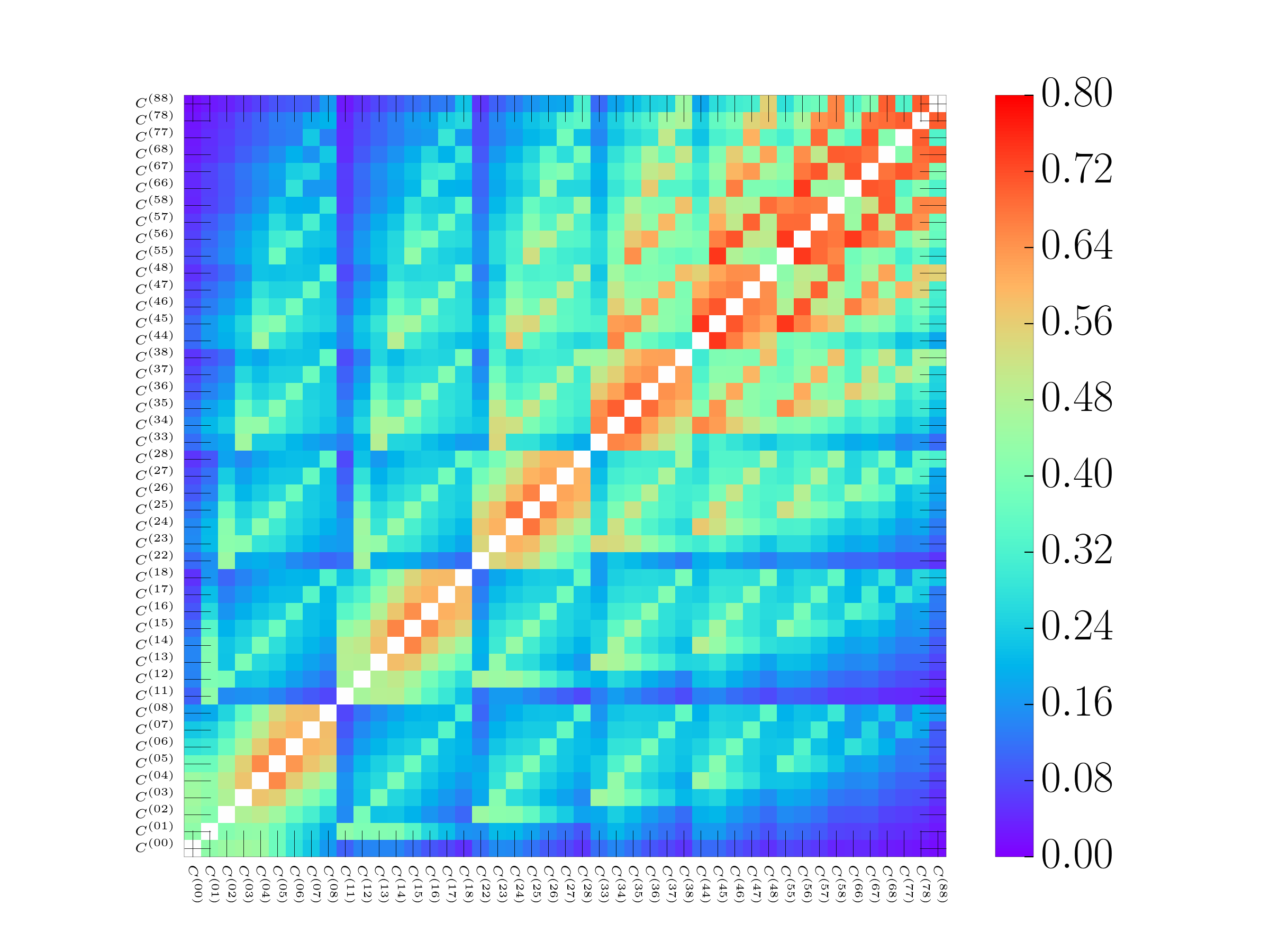}
\includegraphics[trim=70 0 0 0, width=10cm, height=8.5cm]{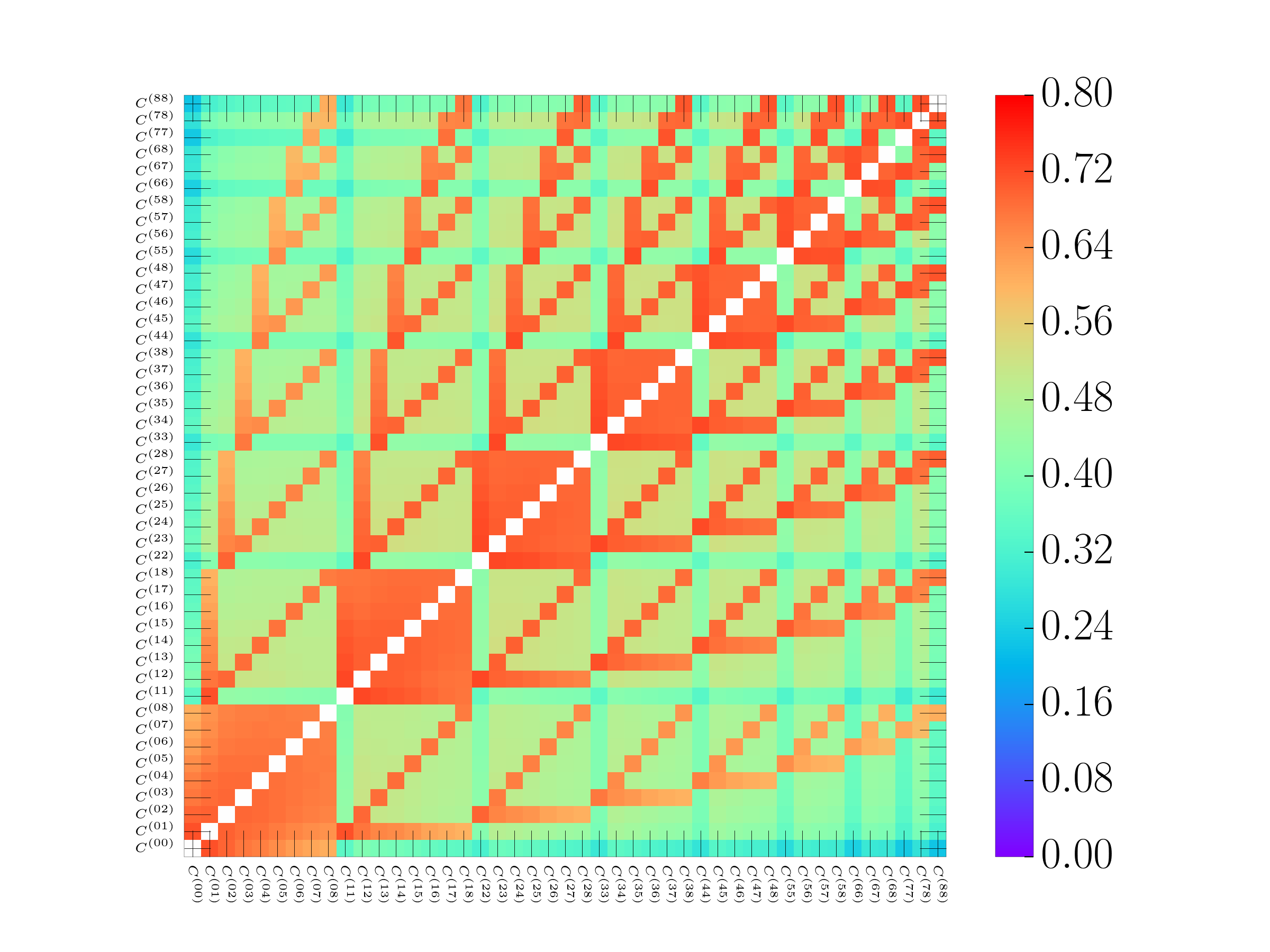}
\caption{The correlation coefficient matrix (45 $\times$ 45) for the power spectra at different multipole bands centered at $\ell=2154$ (left) and $16681$ (right) for SPHEREx.  The $x$- or $y$- axis is labeled by the band power sequence as \{$C^{(0,0)}_{\ell}$, $C^{(0,1)}_{\ell}$, $C^{(0,2)}_{\ell}$, ..., $C^{(8,8)}_{\ell}$\} with 45 power spectra formed among all of the nine SPHEREx broad bands. The superscript ``${ }^{(a,b)}$'' denotes a cross-power spectrum between broad bands $a$ and $b$. The non-negligible correlations in the covariance matrix originate from unique correlation patterns in the covariance matrix of the CIB components, which make the component separation possible. The diagonal elements are masked out in the matrix.}
\label{fig:ccc_spherex}
\end{figure*}

The covariance matrix between band powers $C_{\ell}^{\nu_1\nu_2}$ and bandpower $C_{\ell}^{\nu'_1\nu'_2}$ is given by:
\begin{eqnarray}
\rm{COV}^{(\nu_1\nu_2,\nu'_1\nu'_2)}_{\ell\ell}&=&\langle ({\bf C}_{\ell}^{\nu_1\nu_2}-\langle {\bf C}_{\ell}^{\nu_1\nu_2}\rangle)({\bf C}_{\ell}^{\nu'_1\nu'_2}-\langle {\bf C}_{\ell}^{\nu'_1\nu'_2}\rangle)\rangle\nonumber\\
&=&\frac{1}{(2\ell+1)}[\langle {\bf C}_{\ell}^{\nu_1\nu'_1}\rangle\langle {\bf C}_{\ell}^{\nu_2\nu'_2}\rangle+\langle {\bf C}_{\ell}^{\nu_1\nu'_2}\rangle\langle {\bf C}_{\ell}^{\nu'_1\nu_2}\rangle]\nonumber\\
&=&\frac{1}{(2\ell+1)}[C_{\ell}^{\nu_1\nu'_1}C_{\ell}^{\nu_2\nu'_2}+C_{\ell}^{\nu_1\nu'_2}C_{\ell}^{\nu'_1\nu_2}]\nonumber\\\label{cov0}
\end{eqnarray}
and $\rm{COV}_{\ell\ell'}=0$ if $\ell\neq \ell'$. Here ${\bf C}^{\nu_1\nu_2}_{\ell}$ is a power spectrum ensemble with average being $C^{\nu_1\nu_2}_{\ell}$. The covariance consists of both Gaussian and non-Gaussian components. It is not always true that the non-Gaussian covariance is negligible because the halo sample variance could be significant at small angular scales~\citep{2013MNRAS.429..344K}. In this work, the CIB model is mainly dominated by DGL and shot noise at most of the scales so we only adopt the Gaussian part of the covariance and neglect the non-Gaussian one~\citep{2013arXiv1311.2338D, 1999ApJ...527....1S, 2000ApJ...537....1W, 2001ApJ...554...56C,2010A&A...523A...1J,2014MNRAS.441.2456T}. Based on the covariance matrix, we further define a correlation coefficient matrix at a single multipole $\ell$ as 

\begin{equation}
r^{(\nu_1\nu_2,\nu'_1\nu'_2)}_{\ell\ell}=\frac{{\rm COV}^{(\nu_1\nu_2,\nu'_1\nu'_2)}_{\ell\ell}}{\sqrt{{\rm COV}^{(\nu_1\nu_2,\nu_1\nu_2)}_{\ell\ell}{\rm COV}^{(\nu'_1\nu'_2,\nu'_1\nu'_2)}_{\ell\ell}}}.\label{corr}
\end{equation}

\begin{figure*}
\centering
\includegraphics[width=8cm, height=6cm]{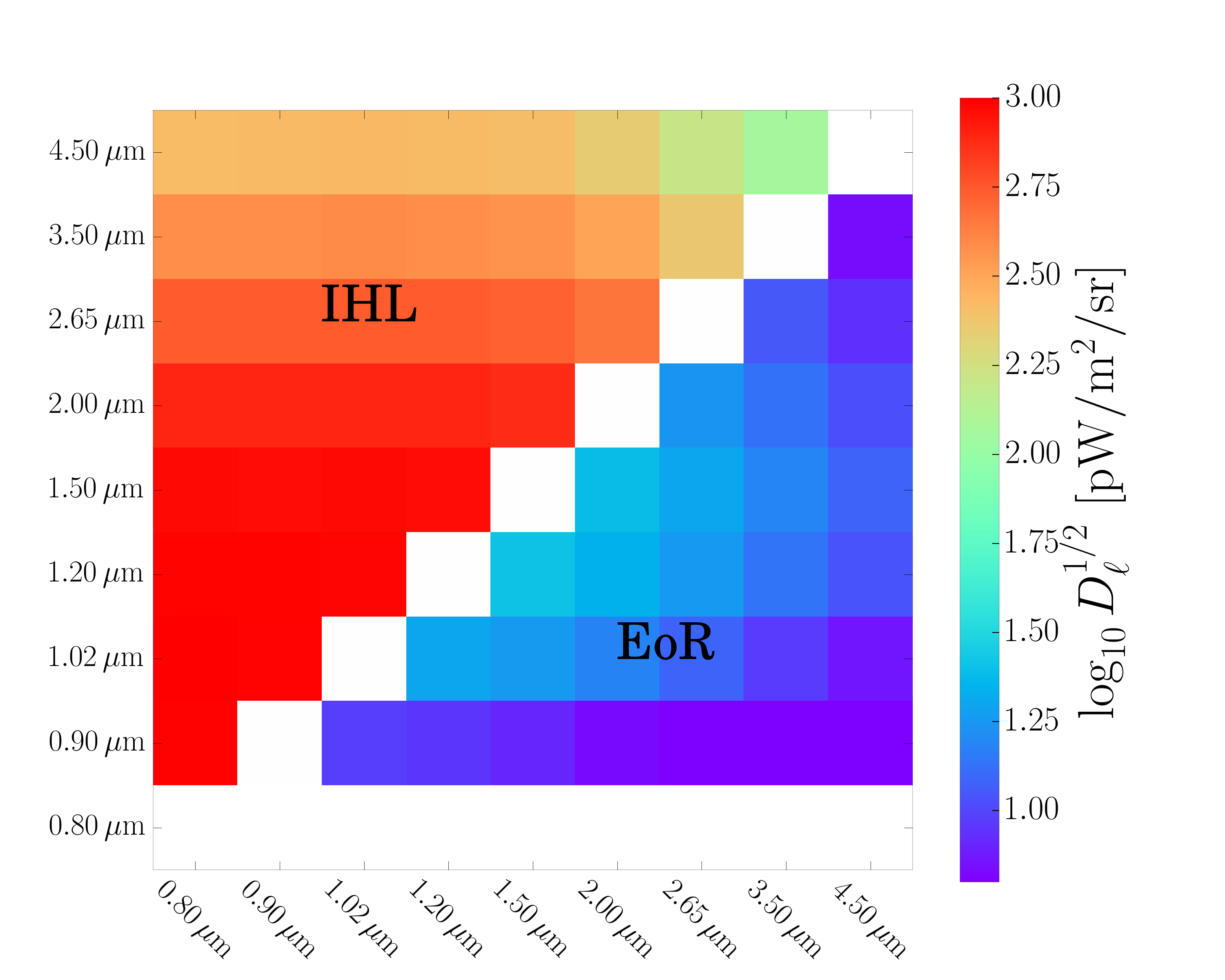}
\includegraphics[width=8cm, height=6cm]{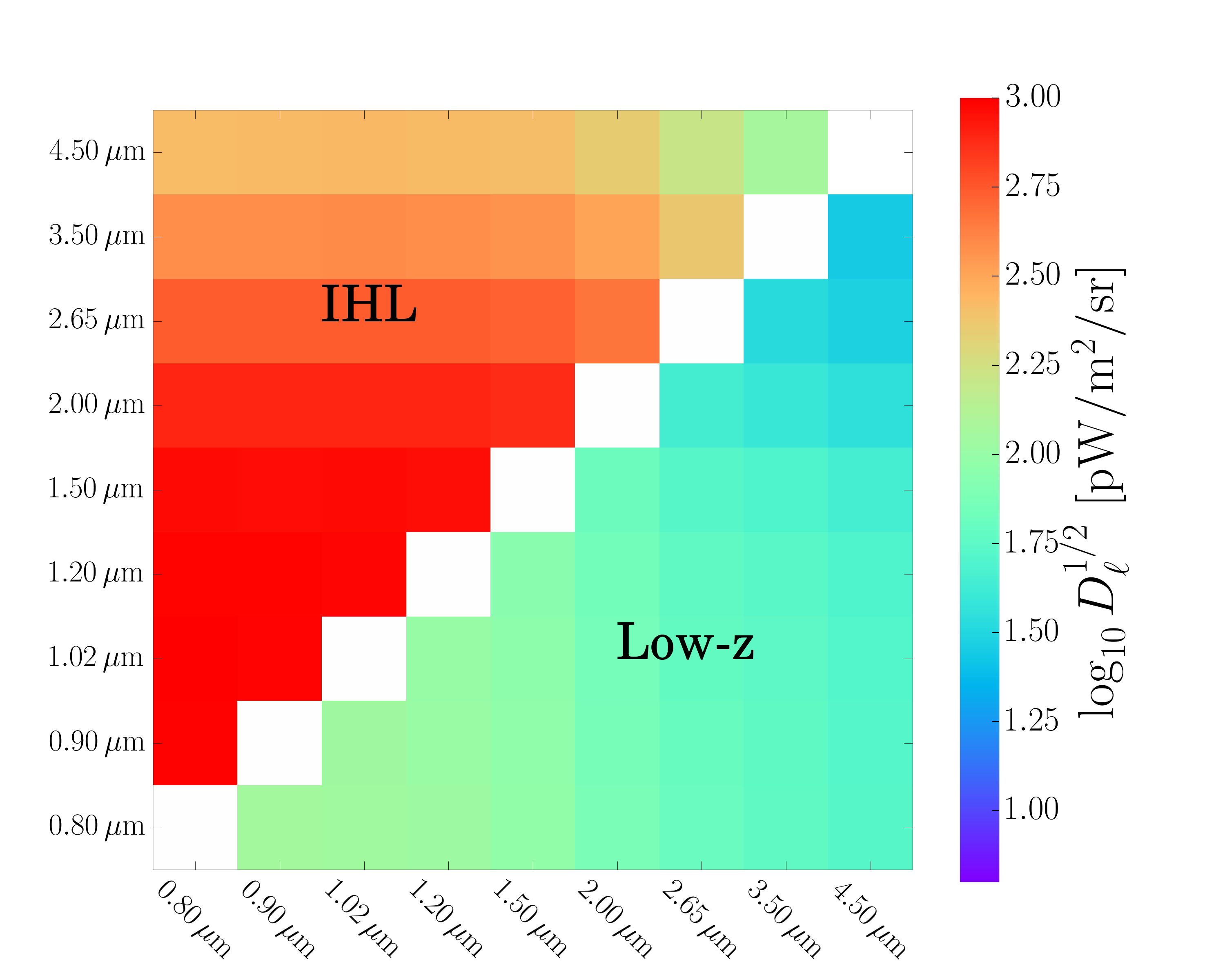}

\includegraphics[width=8cm, height=6cm]{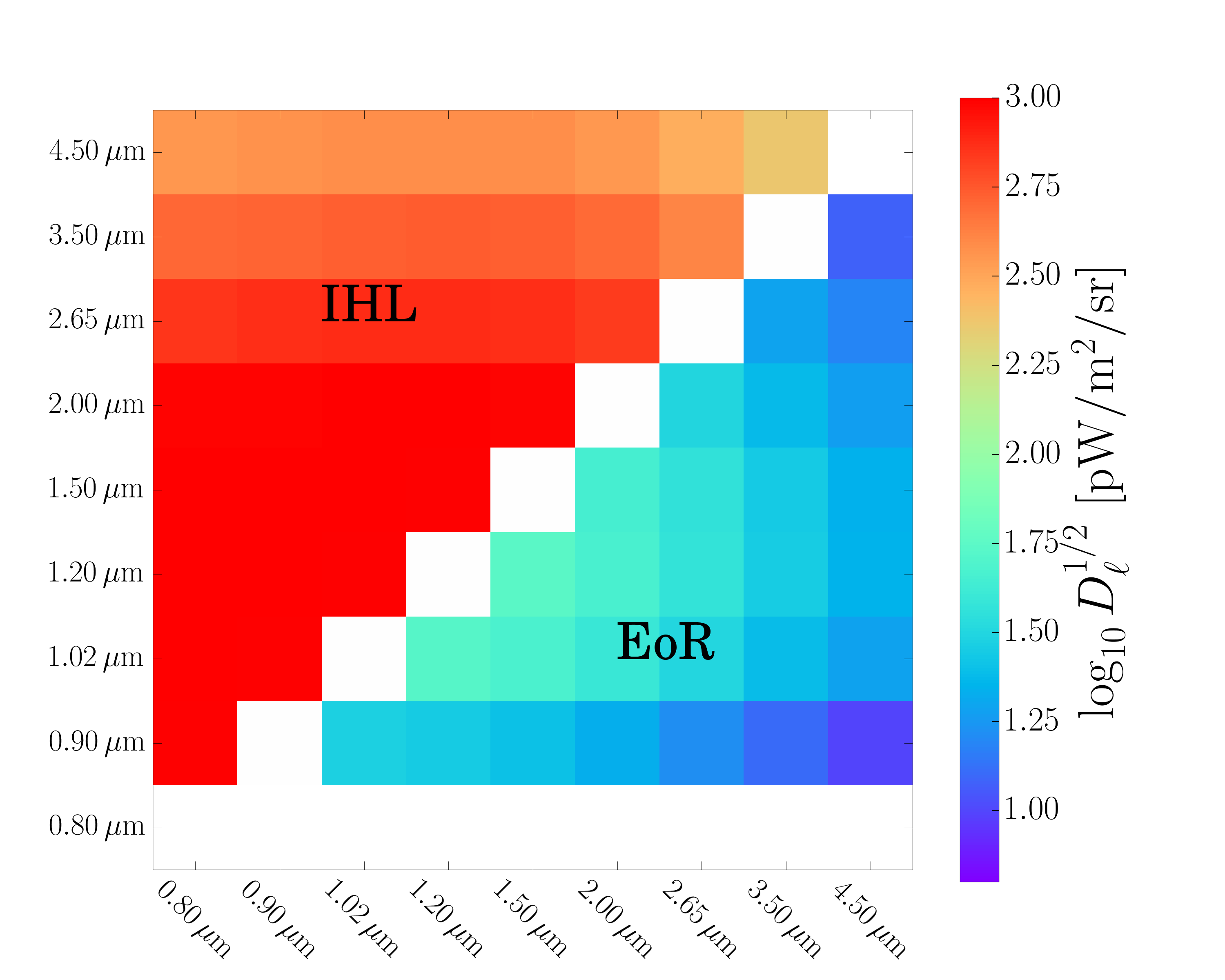}
\includegraphics[width=8cm, height=6cm]{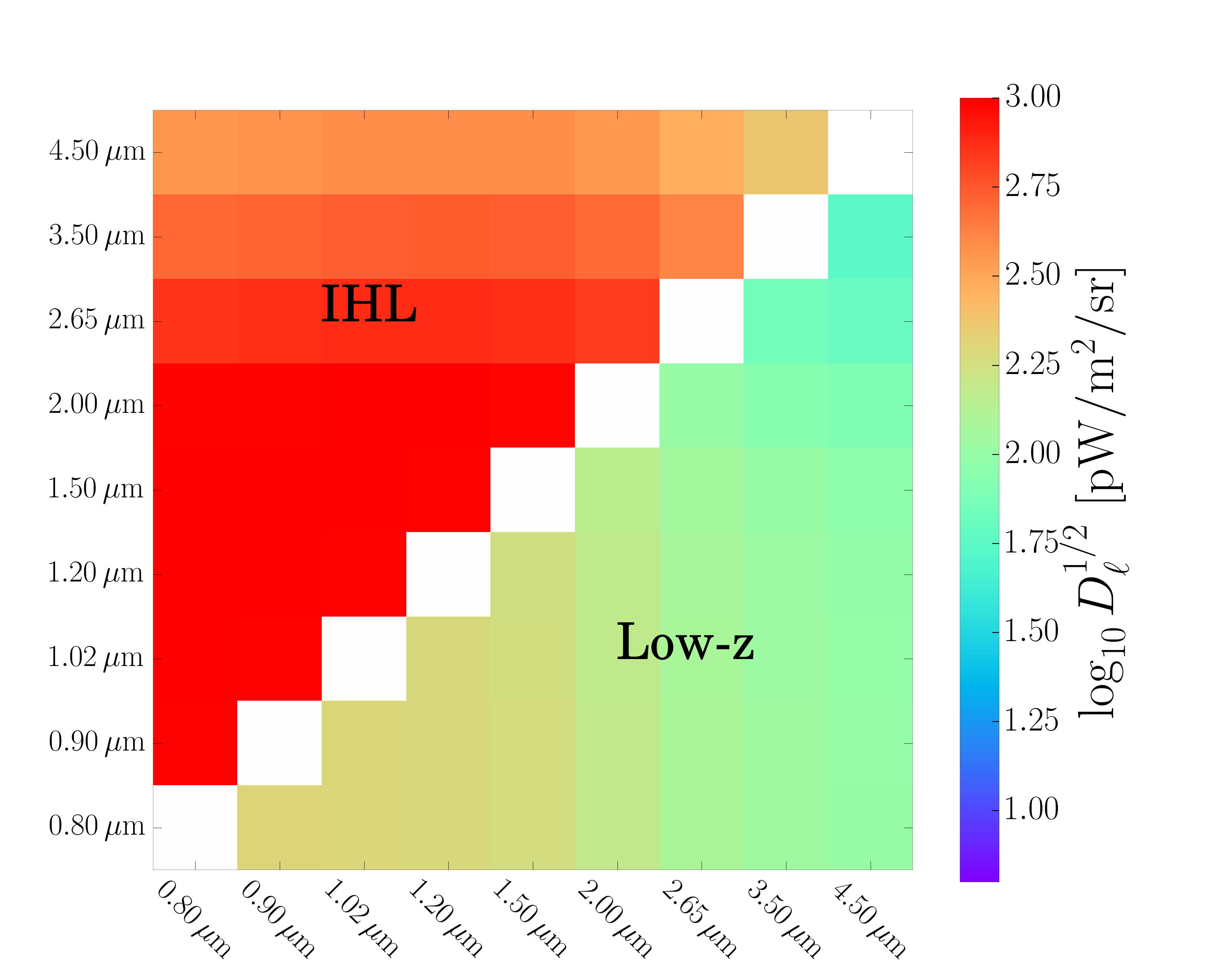}
\caption{The cross-power spectra at $\ell=5000$ (top) and $\ell=50000$ (bottom). The auto power spectra are masked out so the correlation patterns among frequency bands can be easily seen. The white horizontal row in the \eor\ triangle denotes the cross correlations with 0.8 $\um$ at which wavelength the \eor\ signal is null due to the Lyman-$\alpha$ absorption. }\label{corr_component}
\end{figure*}

Realistic power spectrum covariance should take experimental effects into account, such as a partial sky coverage, discrete band powers, width of frequency broad band, and instrumental noise. The modified covariance matrix is given by:
\begin{eqnarray}
{\rm COV}^{(\nu_{p_1}\nu_{p_2},\nu'_{p_1}\nu'_{p_2})}_{\ell_b\ell_b}&=&\frac{1}{(2\ell_b+1)\Delta\ell_b(\epsilon_1\epsilon_2\epsilon'_1\epsilon'_2)^{1/4}(R_{p_1}R_{p_2}R_{p'_1}R_{p'_2})^{1/4}}\nonumber\\&\times&[\tilde C_{\ell_b}^{\nu_{p_1}\nu'_{p_1}}\tilde C_{\ell_b}^{\nu_{p_2}\nu'_{p_2}}
+\tilde C_{\ell_b}^{\nu_{p_1}\nu'_{p_2}}\tilde C_{\ell_b}^{\nu'_{p_1}\nu_{p_2}}],\nonumber\\\label{cov}
\end{eqnarray}
where $\Delta_b$ is the width of the band power $b$, $\nu_p$ is the central value of a broad band averaged within $\nu_1<\nu<\nu_2$ that contains $R_p$ narrow bands, $\epsilon_i$ is the sky fraction for map at $\nu_i$ and the observed power spectrum $\tilde C_{\ell_b}$ = $C_{\ell_b}$ + $N_{\ell_b}$, where $N_{\ell_b}$ is the instrumental noise at band $\ell_b$. There are non-negligible band-band correlations in the full covariance matrix as seen from Fig.~\ref{fig:ccc_spherex}, while they are almost negligible in the covariance matrix constructed from auto-correlations. The non-negligible band-band correlations originate from the unique structures (Fig.~\ref{corr_component}) in the covariance matrix of each component.

This covariance can be reduced to a Knox formula for the power spectrum when $\nu_1$ = $\nu'_1$ and $\nu_2$ = $\nu'_2$, i.e., 
\begin{equation}
\Delta^2(C_{\ell_b}^{\nu_1\nu_2})=\frac{2}{(2\ell_b+1)\Delta\ell_b(\epsilon_1\epsilon_2)^{1/2}}(C_{\ell_b}^{\nu_1\nu_2}+N_{\ell_b}^{\nu_1\nu_2})^2.\label{knox}
\end{equation}
In our following discussion we only adopt a white noise model for simplicity.

The component separation relies on the posterior distribution functions (PDF) which are built on the maximum likelihood:
\begin{eqnarray}
-2\ln\mathcal{L}&=&\displaystyle\sum_{b}\displaystyle\sum_{\substack{\nu_i\nu_j \\ \nu_{i'}\nu_{j'}}}[C^{(\nu_i\nu_j)}_{b}-\hat C^{(\nu_i\nu_j)}_{b}(\textbf{P})]\nonumber\\
&\times&{\rm COV}^{-1,(\nu_i\nu_j),(\nu_{i'}\nu_{j'})}_{bb}\nonumber\\
&\times&[C^{(\nu_{i'}\nu_{j'})}_{b}-\hat C^{(\nu_{i'}\nu_{j'})}_{b}(\textbf{P})]\nonumber\\
&+&2\ln[(2\pi)^{N_p/2}|\textbf{COV}|^{1/2}].\label{like}
\end{eqnarray}
Here $\hat{C_b}$ denotes the theoretical power spectrum, $\nu_i$ is a broad band, $N_p$ is the dimension of the parameter space and we ignore the last term because we assume it is a constant in parameter space~\citep{2009A&A...502..721E,2016MNRAS.456L.132S,2016arXiv160105779K}. It is worth noting that a similar component separation at the cross-frequency angular power spectrum level has been applied to various CMB analyses for the CMB community~\citep{2014A&A...571A..15P, 2012ApJ...746....4M}.\\

\begin{figure}
\includegraphics[width=9cm, height=7.2cm]{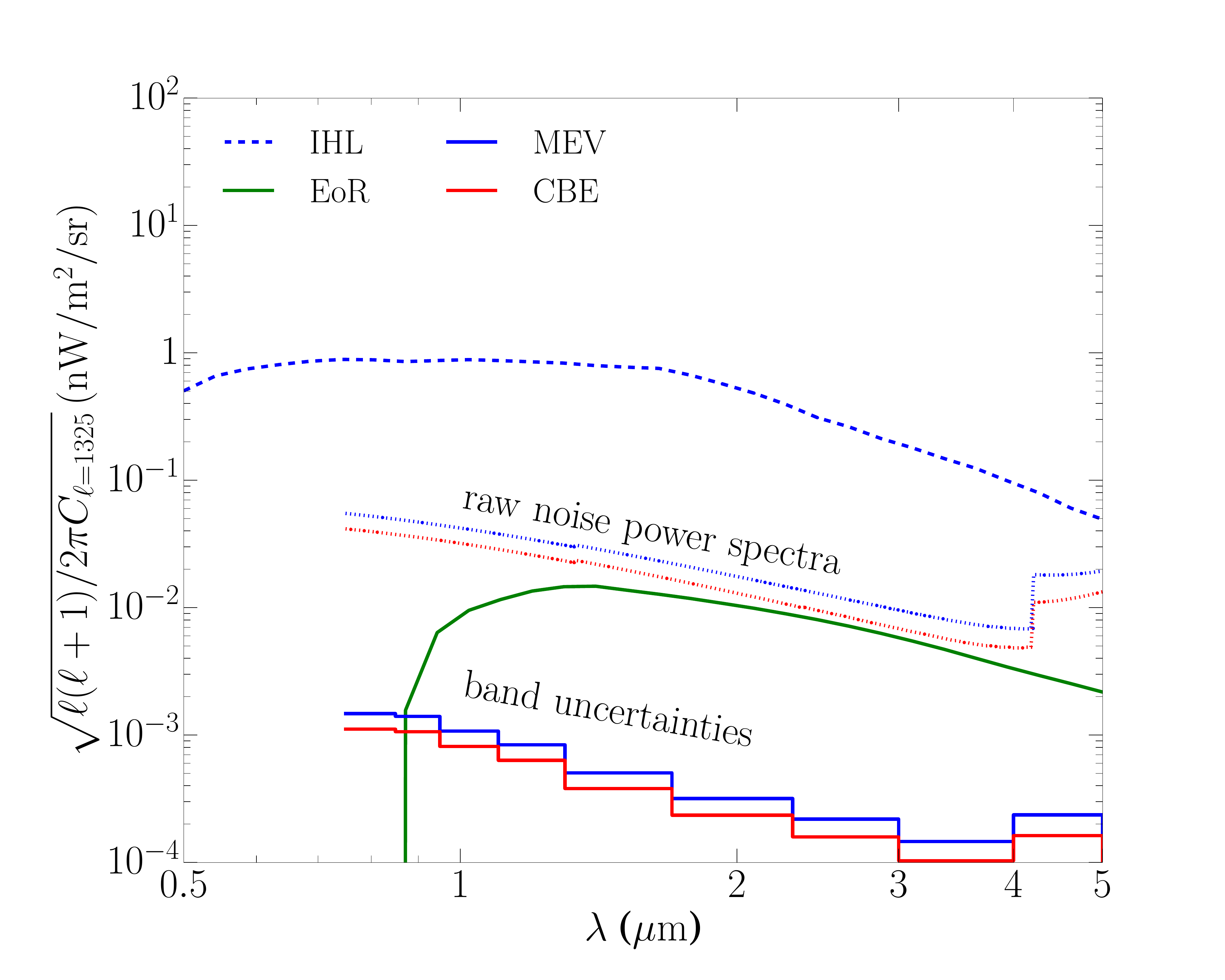}
\caption{The spectrum of large-scale fluctuations of IHL and EoR components for SPHEREx. The dotted lines are the SPHEREx noise power spectra after taking the optimal scan strategy of the deep survey into account. Here the parameters for the \ihl, \lz\,and \eor\, are $A^{\ihl}_i$ =1, $A^{\lz}_{i}$ = 1 and $A^{\eor}_{i}$ = 0.5. The angular scales between $5'$ and $22'$ are used to estimate the band-power uncertainty, and $\ell=1325$ corresponds to the averaged scales. The current best
estimate (CBE) and maximum expected value (MEV) performance
shown as red and blue steps are optimistic and pessimistic
noise levels, respectively~\citep{2018arXiv180505489D}.}\label{spnl}
\end{figure}

\section{SPHEREx forecasting}

Many probes now are proposed to study the intensity mapping of atomic lines and the extragalactic background (\citealt{2013arXiv1305.5422S, 2016arXiv160205178C, 2016SPIE.9904E..0OR}). SPHEREx is a satellite mission aiming at three major scientific goals -- the non-Gaussianity signatures created by inflation, the extragalactic light and interstellar and circumstellar ices. It spans a wavelength range from 0.75 $\um$ to 5.0 $\um$ and has a $6.2''$ spatial resolution. It scans a full sky with two surveys at different depths -- an all-sky shallow survey and two deep surveys near the ecliptic poles. We focus on the deep survey in this work and have constructed a noise model, which is shown in Fig.~\ref{spnl}, from two years scanning on the deep fields. A minimum requirement of 3 \rm{kJy}/\rm{sr} flux sensitivity is imposed on all the bands to maintain a band-power sensitivity of the scientific objective. The current best estimate (CBE) and maximum expected value (MEV) performance shown as red and blue steps are optimistic and pessimistic noise levels, respectively. The optimal scan strategy for a 24-month integration time has been extensively investigated, and we chose the best scanning strategy that results in the most uniform and largest number counts per pixel on the deep fields, to determine effective noise properties at those bands. Different levels of band-power uncertainties have taken the 24-month scan into account, and the dotted lines are noise power spectra. The CBE noise estimation is adopted in this work to forecast the EoR detectability.

\subsection{SPHEREx mock data}
We consider five components that result in CIB spatial fluctuations --- IHL, EoR, \lz\ galaxies, the diffuse Galactic light (DGL) and shot-noise from all the CIB components. The DGL and shot-noise are the dominant components in the CIB fluctuations measured in the nine broad bands of SPHEREx, but their power spectra can be easily modeled as a power-law and a constant, respectively. To determine the amplitude and slope of DGL and shot-noise for SPHEREx, we extrapolate the dependence of DGL amplitude versus wavelength using the model from CIBER's low-resolution spectrometer measurements~\citep{2015ApJ...806...69A}. This model is based on correlations between the near-infrared emission and 100 $\mu$\rm{m} from the infrared astronomical satellite (IRAS) tracing Galactic dust emission, and we normalize the amplitude of the model to CIBER's DGL measurement at 1.1 $\mu$\rm{m}. We extrapolate a relation of shot-noise amplitude versus wavelength from HST auto power spectra at five wavelengths~\citep{2015NatCo...6E7945M} 0.606, 0.775, 0.850, 1.25 and 1.60 $\mu$\rm{m} from the Great Observatories Origins Deep Survey~\citep{2004ApJ...600L..93G, 2011ApJS..193...27W} and normalize it to the CIBER shot-noise level at 1.1 $\mu$\rm{m}. We also adopt the averaged correlation coefficients from CIBER's DGL and HST's shot-noise measurements at different wavelengths. Based on these steps, we generate 45 mock DGL and shot noise components at nine SPHEREx broad bands. For other CIB components, i.e., \ihl, \eor\ and \lz\ galaxies, we use physical models~\citep{2012arXiv1210.6031C,2012ApJ...756...92C,2012ApJ...752..113H} to calculate the power spectra. 

\subsection{Component-separation implementation}
We model uncertainties in the CIB component-power-spectra by considering they are described by a set of known $C^{c,\lambda}_{\ell}$ templates with unknown independent  amplitudes $A^{c}_{\lambda}$ at different wavelengths $\lambda$ (Eq. \ref{component_ps}). Here $c$ refers to \ihl, \eor\ or \lz\ galaxies. We use the standard physical models in the literature to calculate the power spectra of \ihl~~\citep{2012arXiv1210.6031C}, \eor~~\citep{2012ApJ...756...92C} and \lz~~\citep{2012ApJ...752..113H}. The DGL and shot noise are considered known as they can be precisely fitted to the data at large and small angular scales respectively. The total number of parameters for such a minimum parametrization amounts to $\mathcal{O}(3N)$, and $N$ is the number of broad bands. Even with the nine synthesized bands, the component separation scheme becomes computationally challenging as the total number of parameters grows quickly when CIB sources have to be modeled over a large number of broad bands. In this work, the total number of parameters is $3N-1=26$ since the drop-out feature of the SED eliminates one degree of freedom. We note that all the physical parameters of the CIB components as introduced in Section 2, rather than the amplitudes $A^{c}_{\lambda}$, could be varied at each wavelength but the total number of parameters would be an order of magnitude larger than our current parametrization, making computation prohibitive. This parametrization can be viewed as minimum, and we can effectively extract the component information from the measured CIB data.

\begin{figure*}
\centering
\includegraphics[width=17cm, height=13cm]{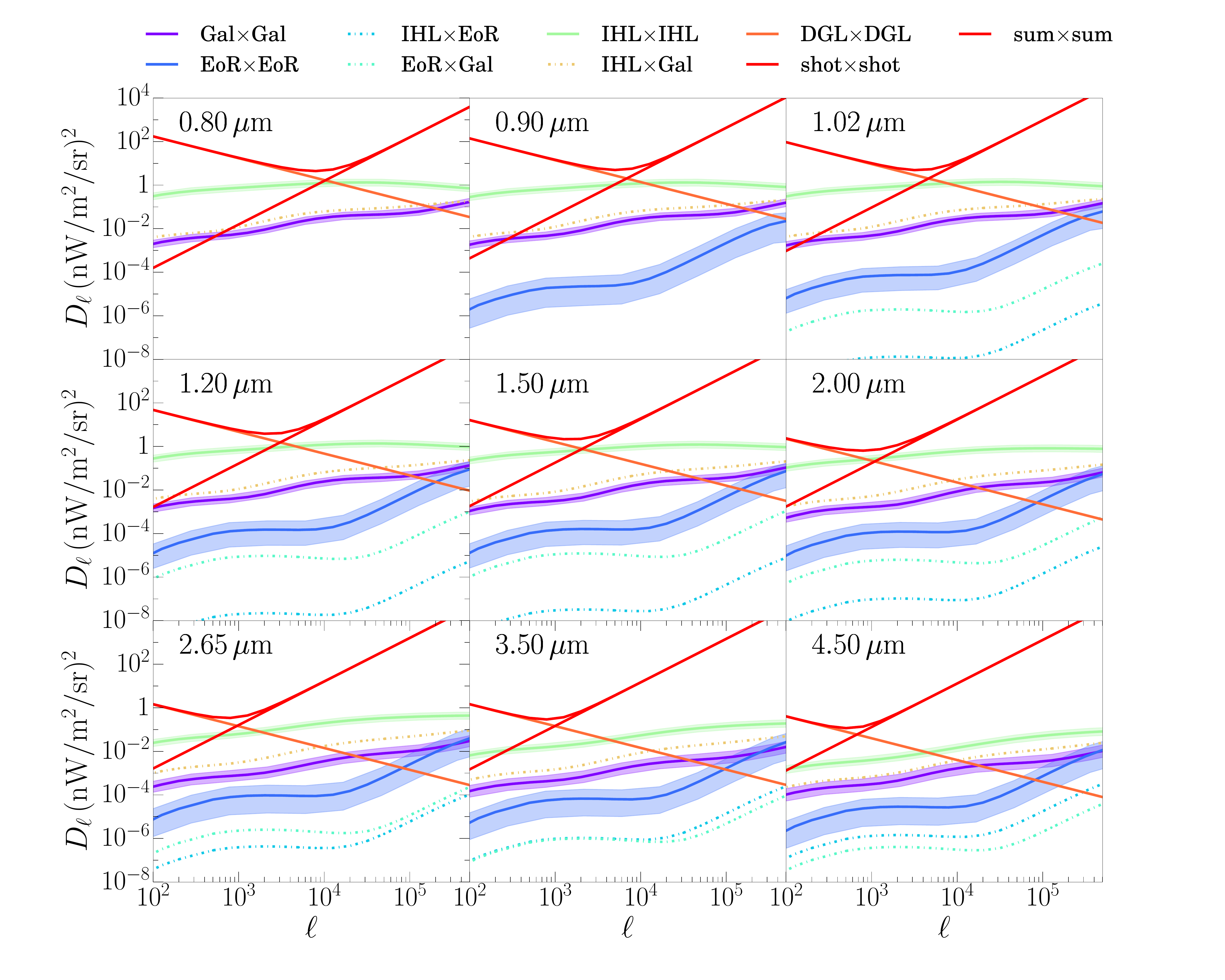}
\caption{Best fit CIB components reconstructed from 45 SPHEREx power spectra made up of 9 auto power spectra in each frequency band and 36 cross-band power spectra. Here, $D_{\ell}=\ell(\ell+1)/(2\pi)C_{\ell}$. In this figure, we only show detailed components for the 9 auto power spectra. In each subplot, for the auto-power spectra (solid lines) of the CIB components \ihl, \lz\ and \eor, we show both the best-fit models and the 1$\sigma$ confidence regions, but only show the best-fit models for the cross-power spectra (dashed lines) among the CIB components. The confidence regions for the CIB components $\ihl$ and $\lz$ in each subplot are enlarged by 40\% for visualization purposes.}\label{spherex_component_all}
\end{figure*}

We create mock CIB power spectra (Fig.~\ref{spherex_component_all}) at all the broad frequency bands with some fiducial parameter set $\{A^{c}_{\lambda}\}$ (Table 2). The power spectra are binned within $10^2<\ell<10^6$ and ten angular bands are made. With all the mock power spectra, we for the first time construct a full covariance matrix 
\begin{equation}
\rm{COV}^{(\lambda_1\lambda_2,\lambda'_1\lambda'_2)}_{\ell\ell}=\langle (C_{\ell}^{\lambda_1\lambda_2}-\bar C_{\ell}^{\lambda_1\lambda_2})(C_{\ell}^{\lambda'_1\lambda'_2}-\bar C_{\ell}^{\lambda'_1\lambda'_2})\rangle,
\end{equation}which is built on the components' auto- and cross-correlations at different wavelengths (Fig.~\ref{spherex_component_all}) and allows us to implement a component separation algorithm. Here $C_{\ell}^{\lambda_1\lambda_2}$ can be a auto-power ($\lambda_1=\lambda_2$) or a cross-power ($\lambda_1\neq\lambda_2$) spectrum for two wavelengths $\lambda_1$ and $\lambda_2$. A correlation coefficient matrix $r^{(\lambda_1\lambda_2,\lambda'_1\lambda'_2)}_{\ell\ell}$ is shown in Fig.~\ref{fig:ccc_spherex} for two different multipole bands. We see that there are non-negligible band-band cross-correlations at most angular scales. Because of the band-band cross correlations and assumptions about the shapes of the $C^{c,\lambda}_{\ell}$, separating each CIB component from the total emission becomes possible. We will further discuss how the component separation would be affected when the power spectra have different shapes between data and simulations.

\begin{figure*}
\includegraphics[trim=140 0 0 0,width=20cm, height=16cm]{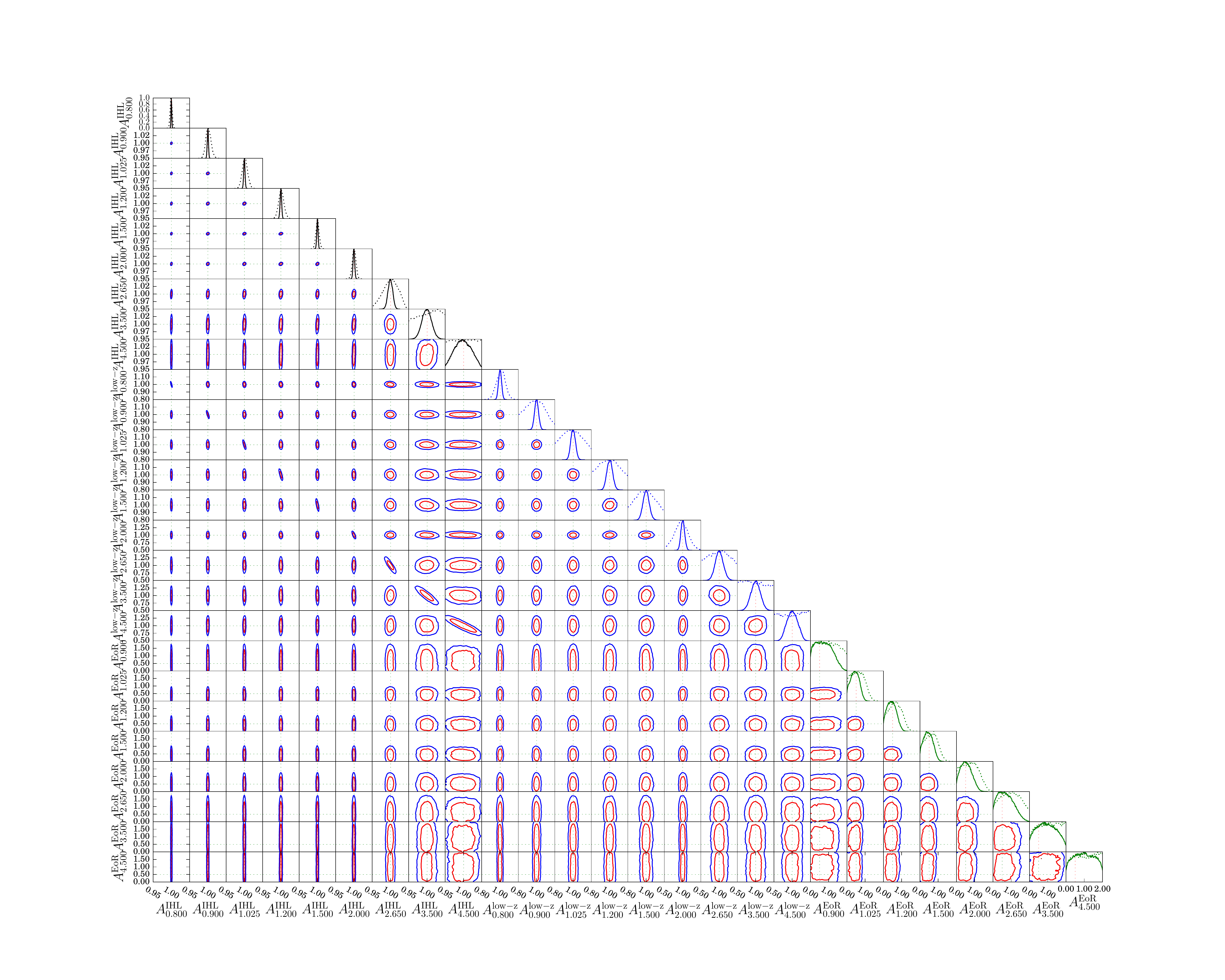}
\caption{The constraints on all the parameters for $\ihl$, $\lz$ and $\eor$ components with all 45 mock SPHEREx power spectra. The dashed vertical and horizontal lines denote the input parameters which are correctly recovered. In diagonal direction, we show 1D likelihood functions for EoR model (a) in dashed lines and EoR model (b) from Table 2 in solid lines. All the 2D contours are plotted for EoR model (b). The improved constraints due to a full covariance can be easily seen from all 1D likelihood functions.}\label{spherex_all}
\end{figure*}

\begin{table*}
\caption{Various $\rm{EoR\,models}$ used for validation.}
\begin{center}\resizebox{\linewidth}{!}{
\begin{tabular}{c|cccccccc} 
 \hline\hline
$\rm{EoR\,models}$&$f_{\rm esc}$&$f_{\ast}$&$A^{\ihl}_{\lambda_i}$&$A^{\lz}_{\lambda_i}$&$A^{\eor}_{\lambda_i}$&$\rm{covariance}$&band range&No. of bands\\
\hline
(a)&0.5&0.03&1&1&0.5&$\rm{9\,auto}$&$0.8\um<\lambda<5\um$&9\\
(b)&0.5&0.03&1&1&0.5&$\rm{45\,auto+cross}$&$0.8\um<\lambda<5\um$&9\\
(c)&0.5&0.03&1&1&1.0&$\rm{45\,auto+cross}$&$0.8\um<\lambda<5\um$&9\\
(d)&0.2&0.06&1&1&0.5&$\rm{45\,auto+cross}$&$0.8\um<\lambda<5\um$&9\\
(e)&0.5&0.03&1&1&0.5&$\rm{45\,auto+cross}$&$0.6\um<\lambda<4\um$&10\\
(f)&0.5&0.03&1&1&1.0&$\rm{45\,auto+cross}$&$0.6\um<\lambda<4\um$&10\\
\hline
\end{tabular}}
\end{center}
\end{table*}

\begin{figure*}
\centering
\includegraphics[width=16cm, height=12cm]{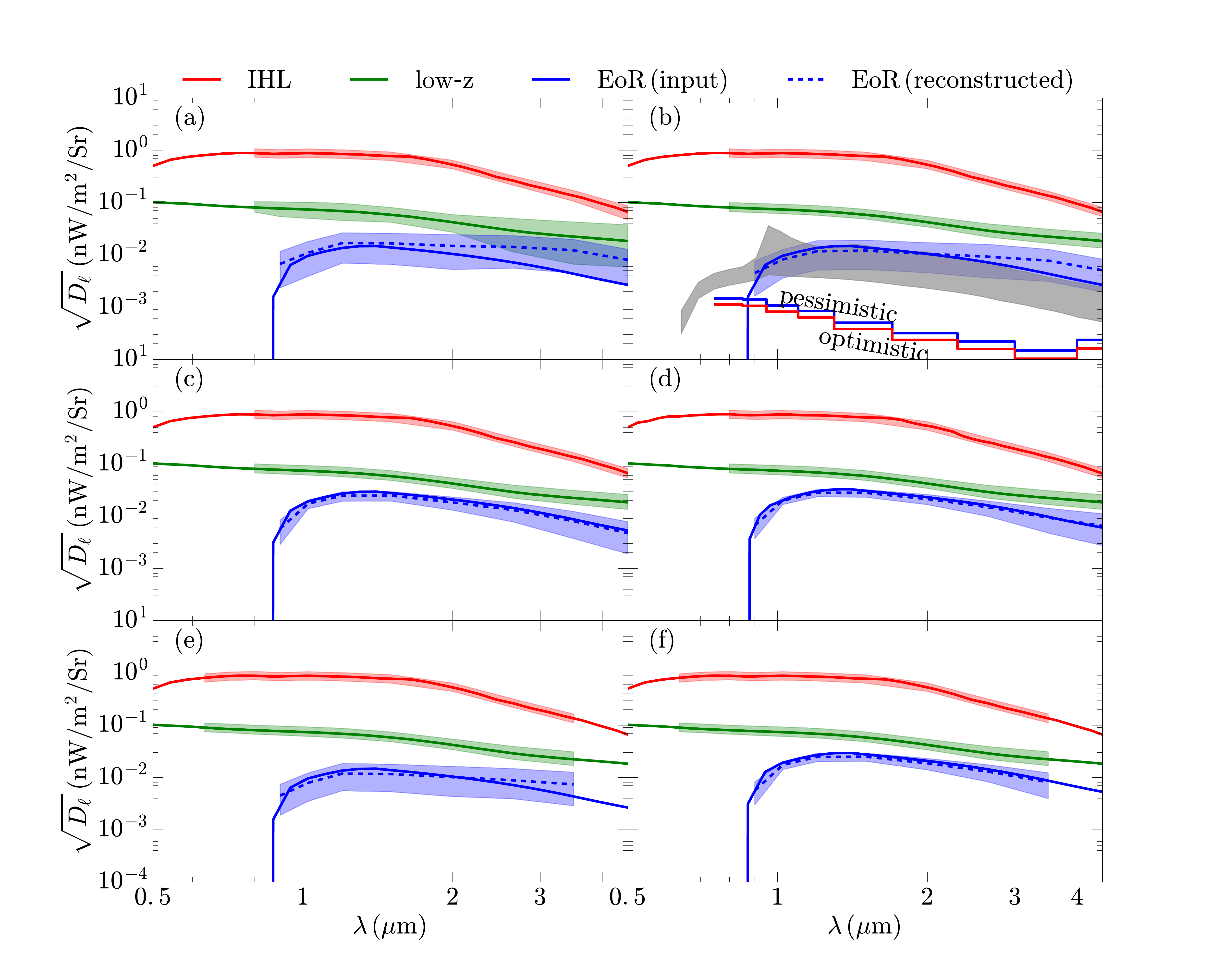}
\caption{The rms anisotropy SED of cosmic infrared background fluctuations versus wavelength. Although being strongly contaminated by the \ihl\ and \lz, the EoR signal can still be detected at $>$ 5$\sigma$ level. In each panel, the blue dashed line is the \eor\ reconstruction from the mock data. Panels (a) to (f) show the reconstructions performed for six EoR models and other parameter variations outlined by cases (a) to (f) in Table 2. Ideally, if the CIB only contains the EoR, it would be detected at $\sim 100\sigma$ based on the optimistic band-power uncertainties. The upper and lower bounds of SPHEREx's \ihl\ and \lz\ components are enlarged by 20\% to be visible. All the amplitudes are taken from the band power centered at $\ell=1325$.}\label{sed}
\end{figure*}

\begin{figure*}
\centering
\includegraphics[width=16cm, height=12cm]{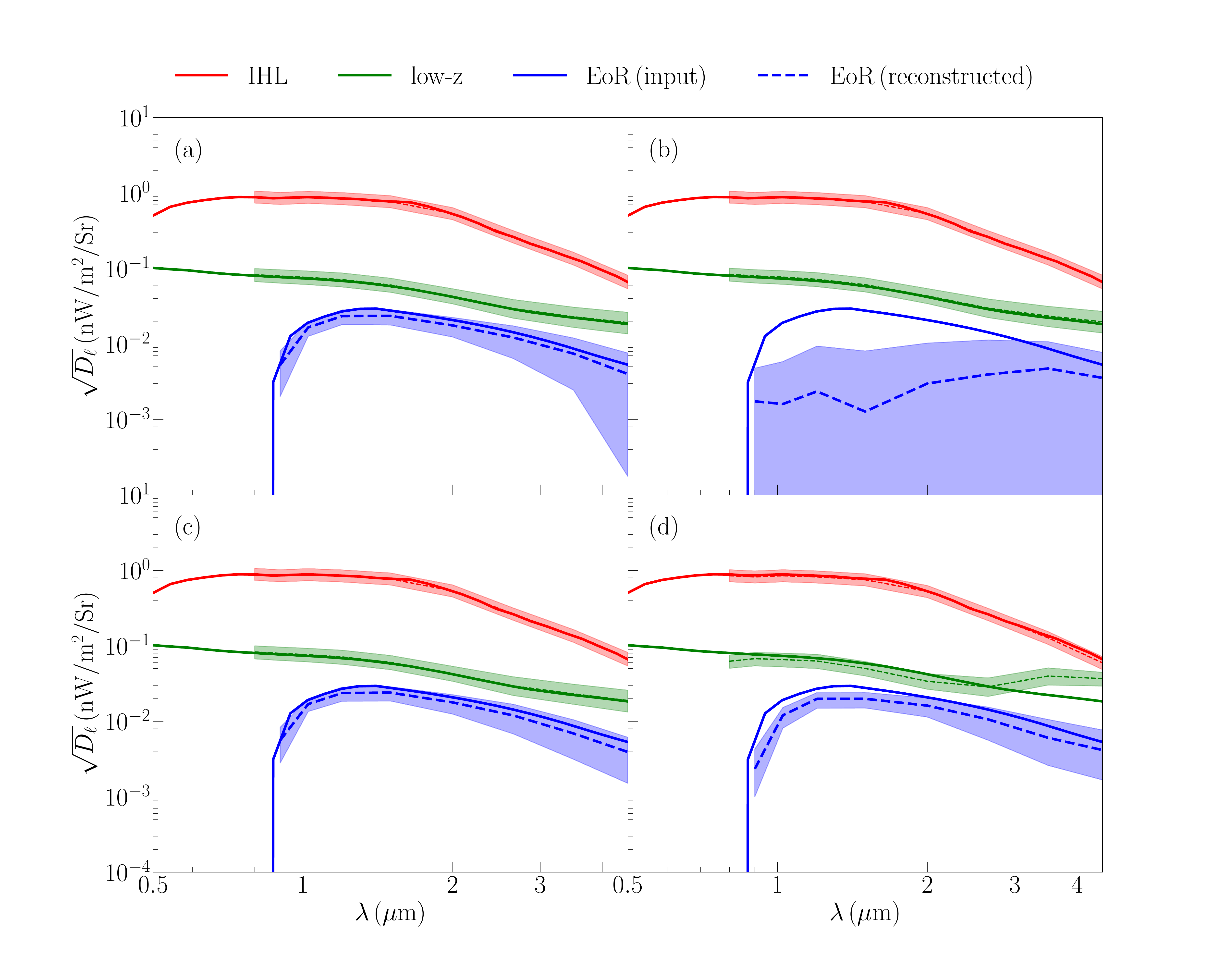}
\caption{Various tests for the reconstructed SEDs with different assumptions. (a) The amplitudes $A^{\eor}_{\lambda}$ can be negative; (b) the \ihl\ power spectra $C^{\ihl}_{\ell}$  are replaced by $C^{\ihl}_{\ell}+2C^{\eor}_{\ell}$ to make the \ihl\ component-covariance correlated with the \eor; all the \ihl, \lz, and \eor\ power spectra are slightly distorted in frequency-space by $(\lambda/(1 \um))^{-0.2}$ (c) and $\ell$-space by $1.0+0.1\log_{10}{(\ell/10^3)}$ (d), respectively. In each panel, the dashed lines are the component-reconstructions from the mock data. The upper and lower bounds of SPHEREx's \ihl\ and \lz\ components are enlarged by 20\% to be visible.}\label{sedtests}
\end{figure*}

We create simulated data for IHL, $\lz$ galaxies and EoR components with parameter sets $\{\hat A^{c}_{\lambda}\}$ sampled by Markov chain Monte Carlo (MCMC) and the component separation procedure is thus established with a maximum likelihood function that involves the mock data, the simulated data and the full covariance. Using MCMC samples, we find that all the posterior distribution functions (PDFs) of IHL, $\lz$ and EoR components peak at the input models. Fig. \ref{spherex_all} shows this component separation can recover both the amplitudes and the spectral SED shapes of IHL, $\lz$, and EoR components, even if the IHL power is four orders of magnitude higher than the EoR. To see how much the full covariance scheme can improve the overall signal-to-noise ratio of EoR signature, we also run simulations with the covariance matrix that is only constructed by auto-power spectra. All posterior PDFs are found to have wider confidence regions, as compared to Fig. \ref{spherex_all}, and all the 1$\sigma$ band-power uncertainties at all the nine wavelengths are much larger than those in Fig.~\ref{spherex_component_all}. 

The 1$\sigma$ uncertainties of the component auto-power spectra for \ihl, \lz\ and \eor\ at nine wavelengths are also estimated from these simulated data (Fig.~\ref{spherex_component_all}). Furthermore, the reconstructed SEDs for the components are obtained from these power spectra at different wavelengths, and the EoR detection significance against the theoretical model can be estimated for the SPHEREx project (Fig.~\ref{sed}).

\subsection{Various tests for the component separation}
We run simulations of an alternative EoR model with different escape fraction $f_{\rm esc}=0.2$ and star formation efficiency $f_{\ast}=0.06$. The EoR component is successfully recovered with a significance around 10$\sigma$ for noise and observational properties consistent with SPHEREx deep fields (Fig.~\ref{sed} (d)). In addition, we test scaling relation between the EoR detection significance and the input amplitude of the EoR model. We increase the amplitude of the EoR model by a factor of two and re-run the component separation. The result verifies that there is a linear scaling relation between the input amplitude and detection significance (Fig.~\ref{sed} (c)). We also revisit the EoR model with no Lyman alpha cutoff~\citep{2014arXiv1412.4872D}, in which a characteristic sharp bump around $1\mu\rm{m}$ is seen. We create both simulations and mock data using a model with the characteristic shape like the gray region in Fig. \ref{sed} (b). Repeating the same procedures, the input SED that contains a sharp bump is reconstructed. 

To investigate how the EoR detection significance could be varied by a different number of broad bands, we include two extra broad bands -- 0.6 $\mu\rm{m}<\lambda < 0.67 \mu\rm{m}$ and 0.67 $\mu\rm{m} < \lambda < 0.75 \mu\rm{m}$ and exclude the 4 $\mu\rm{m}<\lambda<5\mu\rm{m}$ because the instrumental noise at this band is much higher (Fig.~\ref{spnl}) so that it negligibly contributes to the overall signal-to-noise ratio. For this extended model that consists of 10 broad bands, all the amplitudes and shapes are well reconstructed and the resulting SED (Fig.~\ref{sed} (e)) indicates a $\sim 5\sigma$ detection significance, while a higher amplitude EoR model can almost linearly boost the detection significance by a factor of two (Fig.~\ref{sed} (f)). This test means that adding more bands below the Lyman absorption is not necessary.

To investigate how galaxy masking would affect component separation, we create mock data with no masking of bright galaxies. The amplitudes and shapes of the input models are still correctly recovered, although now the \lz\ component is almost two orders of magnitude brighter than the one with certain masking. We conclude deep masking is not required by this component separation scheme. 

We introduce a small random perturbation to the covariance associated with the IHL component, mimicking IHL modeling uncertainties. The numerical tests show a very similar plot to Fig. \ref{sed} (b). Moreover, we create mock EoR band-powers with model parameters ($f_{\rm esc}, f_{\ast}$) = (0.2, 0.06) but make simulations and covariance matrix with model parameters ($f_{\rm esc}, f_{\ast}$) = (0.5, 0.03) to perform a similar test for EoR modeling errors. Although the simulations are performed with a different theoretical model, the reconstructed SED is still identical to that of the mock data. From both tests, we find that the component separation has a great tolerance for model uncertainties, and a perfect knowledge of the models is not strongly required. However, under some extreme conditions, the component separation may fail and we will discuss it in the following text.

To further validate this component separation method, we perform an array of additional tests. The component separation might be prior driven and we test this by allowing the amplitudes $A^{\eor}_{\lambda}$ to be negative. Specifically, the fiducial value $A^{\eor}_{\lambda}$ is 1 and the flat prior is set to $-0.5<A^{\eor}_{\lambda}<2$. The test in Fig.~\ref{sedtests} (a) shows that an extended prior range can negligibly affect the reconstructions. Also, the component separation may fail if there are possible correlations among the CIB components. To check this point, we replace all the \ihl\ power spectra $C^{\ihl}_{\ell}$  by $C^{\ihl}_{\ell}+2C^{\eor}_{\ell}$ so the \ihl\ component-covariance is correlated with the \eor. We find that the reconstructed \eor\ amplitude in such an extreme case is consistent with zero (Fig.~\ref{sedtests} (b)), but the rest are well recovered. Moreover, given the fact that the power-spectrum templates from the theoretical models may differ from the true ones in the data, we investigate how theory differences between data and simulations can affect the component separation, besides those discussed earlier. To test this, we slightly distort all the \ihl, \lz, and \eor\ power spectra in frequency-space by $(\lambda/(1 \um))^{-0.2}$. Fig.~\ref{sedtests} (c) shows that the component separation is not affected by such a distortion in the simulated CIB data that have different SED shapes. Moreover, we multiply all the power spectra in $\ell$-space by an ad hoc function $1.0+0.1\log_{10}{(\ell/10^3)}$, mimicking distortions in the angular scales. The power spectrum differences produced by such a distortion function in $\ell$-space are significant. For example, the deviation of  \ihl\ power spectrum from the original is as large as $\sim100\sigma$. Applying a significant deviation in angular space between the simulations and mock data, unsurprisingly, we find that the \lz\ and \eor\ reconstructions are biased in Fig.~\ref{sedtests} (d).
With high sensitivity CIB power-spectrum measurements at multiple broad bands in the future, there is a possibility that this component separation can be even applied to individual $\ell$-bins at multiple wavelengths, thus the $\ell$-space amplitude and shape will be optimally reconstructed. We defer this discussion to future work.

\section{conclusions}
In this paper, we have calculated all the covariant power spectra among the $\ihl$, $\lz$ and $\eor$ components using the halo-model formalism. We create consistent DGL and shot noise models for SPHEREx and generate $\ihl$, $\lz$ and $\eor$ models from theoretical predictions. Mock CIB auto- and cross-power spectra at nine broad bands \mbox{--} 0.8, 0.9, 1.025, 1.2, 1.5, 2.0, 2.65, 3.5, 4.5 $\mu$\rm{m} are made, and the component separation method is applied to separate EoR signal from IHL and other signals with simulated data expected from 
the planned SPHEREx project. Using simulations, we find that the component separation procedure constructed by a maximum likelihood function with a full covariance among different wavelengths can successfully reconstruct any component in the CIB fluctuations without introducing any significant biases into each reconstructed component. From parameter samples of the Monte Carlo Markov chain, power-spectrum uncertainties are determined for all the auto-power spectra of $\ihl$, $\lz$ and EoR. The spectrum of EoR fluctuations measured by SPHEREx is constructed at a broad range of wavelengths, indicating that a $>5\sigma$ level detection significance can be reached for a wide range of input models and assumptions about the measurement. The component separation algorithm also passed a series of tests related to model assumptions and observational parameter changes such as the number and location of observing bands, verifying that it is generally robust to such variations.

\newpage

\section{Acknowledgments}
We thank the SPHEREx team for insightful comments that improved the quality of the paper. Part of the research was carried out at the Jet Propulsion Laboratory, California Institute of Technology, under a contract with the National Aeronautics and Space Administration. 
CF acknowledges support from NASA grants NASA
NNX16AJ69G, NASA NNX16AF39G, Ax Foundation
for Cosmology at UC San Diego, and the Brand and Monica Fortner Chair. MBS acknowledges the Netherlands Foundation for Scientific Research support through the VICI grant 639.043.006. MGS acknowledges support from the South African Square Kilometre Array Project and National Research Foundation (Grant No. 84156).

\bibliography{ihl_eor}
\end{document}